\documentclass[sigconf]{acmart}

\usepackage{algorithmic}
\usepackage{graphicx}
\usepackage{graphics}
\usepackage{textcomp}
\usepackage{xcolor}

\usepackage[disable]{todonotes}
\usepackage{tikz}
\usepackage{listings}
\usepackage[english]{babel}
\usepackage{pgfplots}
\usepackage{float}
\usepackage{amsthm}
\usepackage[normalem]{ulem}
\usepackage{appendix}
\usepackage{algorithm}
\usepackage{pgfplotstable}
\usepackage{pifont}
\usepackage{placeins}
\usepackage{cleveref}
\usepackage{float}
\usepackage{orcidlink}

\usepgfplotslibrary{fillbetween}
\usepgfplotslibrary{statistics}
\usetikzlibrary{patterns}

\newenvironment{myindentpar}[1]{\begin{list}{}{\setlength{\leftmargin}{#1}}\item[]}
{\end{list}}

\usepackage[mode=buildnew]{standalone}\usepackage{filecontents}

\lstdefinelanguage{riscv}
{
	alsoletter={.},
	alsodigit={0x},
	morekeywords=[1]{
		lb,
lh,
lw,
lbu,
lhu,
		sb,
sh,
sw,
		sll,
slli,
srl,
srli,
sra,
srai,
		add,
addi,
sub,
lui,
auipc,
		xor,
xori,
or,
ori,
and,
andi,
		slt,
slti,
sltu,
sltiu,
		beq,
bne,
blt,
bge,
bltu,
bgeu,
		j,
jr,
jal,
jalr,
ret,
		scall,
break,
nop,
		mul,
bb,
li,
lcnt
	},
	morekeywords=[2]{
		.align,
.ascii,
.asciiz,
.byte,
.data,
.double,
.extern,
		.float,
.globl,
.half,
.kdata,
.ktext,
.set,
.space,
.text,
.word
	},
	morekeywords=[3]{
		zero,
ra,
sp,
gp,
tp,
s0,
fp,
		t0,
t1,
t2,
t3,
t4,
t5,
t6,
		s1,
s2,
s3,
s4,
s5,
s6,
s7,
s8,
s9,
s10,
s11,
		a0,
a1,
a2,
a3,
a4,
a5,
a6,
a7,
		ft0,
ft1,
ft2,
ft3,
ft4,
ft5,
ft6,
ft7,
		fs0,
fs1,
fs2,
fs3,
fs4,
fs5,
fs6,
fs7,
fs8,
fs9,
fs10,
fs11,
		fa0,
fa1,
fa2,
fa3,
fa4,
fa5,
fa6,
fa7,
		ls1
	},
	basicstyle=\footnotesize\ttfamily,
	showstringspaces=false,
	breaklines=true,
	keywordstyle=[1]\color{blue!80!black},
	keywordstyle=[2]\color{orange!80!black},
	keywordstyle=[3]\color{red!50!black},
	stringstyle=\color{mauve},
	commentstyle=\itshape\color{green!50!black},
	morecomment=[l]{;},
	morecomment=[l]{\#},
	morestring=[b]",
	morestring=[b]'
}

\definecolor{mauve}{rgb}{0.58,0,0.82}

\setcopyright{none} \acmConference[ArXiv Version]{}{2021}{May}
\acmYear{2021}

\begin{document}
\title{BasicBlocker: ISA Redesign to Make Spectre-Immune CPUs Faster} 

\author{Jan Philipp Thoma \orcidlink{0000-0003-1613-732X}}
\email{jan.thoma@rub.de}
\orcid{0000-0003-1613-732X}
\affiliation{\institution{Horst-Görtz Institute\\Ruhr-University Bochum}
\streetaddress{}
\city{Bochum}
\country{Germany}
}

\author{Jakob Feldtkeller \orcidlink{0000-0001-9797-1257}}
\email{jakob.feldtkeller@rub.de}
\orcid{0000-0001-9797-1257}
\affiliation{\institution{Horst-Görtz Institute\\Ruhr-University Bochum}
\streetaddress{}
\city{Bochum}
\country{Germany}
}

\author{Markus Krausz \orcidlink{0000-0002-1362-423X}}
\email{markus.krausz@rub.de}
\orcid{0000-0002-1362-423X}
\affiliation{\institution{Horst-Görtz Institute\\Ruhr-University Bochum}
\streetaddress{}
\city{Bochum}
\country{Germany}
}

\author{Tim Güneysu \orcidlink{0000-0002-3293-4989}}
\email{tim.gueneysu@rub.de}
\orcid{0000-0002-3293-4989}
\affiliation{\institution{Horst-Görtz Institute\\Ruhr-University Bochum}
\streetaddress{}
\city{Bochum}
\country{Germany}
}
\affiliation{\institution{DFKI GmbH, Cyber-Physical Systems}
\streetaddress{}
\city{Bremen}
\country{Germany}
}

\author{Daniel J. Bernstein}
\email{djb@cr.yp.to}
\affiliation{\institution{Horst-Görtz Institute\\Ruhr-University Bochum}
\streetaddress{}
\city{Bochum}
\country{Germany}
}
\affiliation{\institution{University of Illinois at Chicago}
\streetaddress{}
\city{Chicago}
\country{USA}
}

\renewcommand{\shortauthors}{J. Thoma, J. Feldtkeller, M. Krausz, T. Güneysu, D. J. Bernstein}

\begin{abstract}
Recent
research
has
revealed
an
ever-growing
class
of
microarchitectural
attacks
that
exploit
speculative
execution,
a
standard
feature
in
modern
processors.
Proposed
and
deployed
countermeasures
involve
a
variety
of
compiler
updates,
firmware
updates,
and
hardware
updates.
None
of
the
deployed
countermeasures
have
convincing
security
arguments,
and
many
of
them
have
already
been
broken.

The
obvious
way
to
simplify
the
analysis
of
speculative-execution
attacks
is
to
eliminate
speculative
execution.
This
is
normally
dismissed
as
being
unacceptably
expensive,
but
the
underlying
cost
analyses
consider
only
software
written
for
current
instruction-set
architectures,
so
they
do
not
rule
out
the
possibility
of
a
new
instruction-set
architecture
providing
acceptable
performance
without
speculative
execution.
A
new
ISA
requires
compiler
and
hardware
updates,
but
these
are
happening
in
any
case.

This
paper
introduces
\textit{BasicBlocker},
a
generic
ISA
modification
that
works
for
all
common
ISAs
and
that
allows
non-speculative
CPUs
to
obtain
most
of
the
performance
benefit
that
would
have
been
provided
by
speculative
execution.
To
demonstrate
the
feasibility
of
BasicBlocker,
this
paper
defines
a
variant
of
the
RISC-V
ISA
called
BBRISC-V
and
provides
a
thorough
evaluation
on
both
a
5-stage
in-order
soft
core
and
a
superscalar
out-of-order
processor
using
an
associated
compiler
and
a
variety
of
benchmark
programs.

\end{abstract}

\maketitle

\section{Introduction}
\label{sec:intro}

The
IBM
Stretch
computer
in
1961
automatically
speculated
that
a
conditional
branch
would
not
be
taken:
it
began
executing
instructions
after
the
conditional
branch,
and
rolled
the
instructions
back
if
it
turned
out
that
the
conditional
branch
was
taken.
More
sophisticated
branch
predictors
appeared
in
several
CPUs
in
the
1980s,
and
in
Intel's
first
Pentium
CPU
in
1993.

Software
analyses
in
the
1980s
such
as~\cite{1982/clark}
reported
that
programs
branched
every
4--6
instructions.
Each
branch
needed
3
extra
cycles
on
the
Pentium,
a
significant
cost
on
top
of
4--6
instructions,
especially
given
that
the
Pentium
could
often
execute
2
instructions
per
cycle.
However,
speculative
execution
removed
this
cost
whenever
the
branch
was
predicted
correctly.

Subsequent
Intel
CPUs
split
instructions
into
more
pipeline
stages
to
support
out-of-order
execution
and
to
allow
higher
clock
speeds.
The
penalty
for
mispredictions
grew
past
10
cycles.
Meanwhile
the
average
number
of
instructions
per
cycle
grew
past
2,
so
the
cost
of
each
mispredicted
branch
was
more
than
20
instructions.
Intel
further
improved
its
branch
predictors
to
reduce
the
frequency
of
mispredictions;
see~\cite{2020/fog}.

Today
the
performance
argument
for
branch
prediction
is
standard
textbook
material.
Accurate
branch
predictors
are
normally
described
as
``critical''
for
performance,
``essential'',
etc.;
see,
e.g.,~\cite{DBLP:conf/isca/CalderG94,DBLP:conf/isca/JuanSN98,2010/gwennap}.
Deployed
CPUs
vary
in
pipeline
lengths,
but
speculative
execution
is
common
even
on
tiny
CPUs
with
just
a
few
pipeline
stages,
and
is
universal
on
larger
CPUs.

This
pleasant
story
of
performance
improvements
was
then
rudely
interrupted
by
Spectre
\cite{kocher2019spectre},
which
exploited
speculative
behavior
in
various
state-of-the-art
CPUs
to
bypass
critical
security
mechanisms
such
as
memory
protection,
stealing
confidential
information
via
hardware-specific
footprints
left
by
speculatively
executed
instructions.
This
kicked
off
an
avalanche
of
emergency
software
security
patches,
firmware
updates,
CPU
modifications,
papers
proposing
additional
countermeasures
targeting
various
software
and
hardware
components
in
the
execution
flow
with
an
impact
on
performance,
while
still
papers
appear
presenting
new
attacks.
Some
countermeasures
have
been
already
broken,
and
it
is
difficult
to
analyze
whether
the
unbroken
countermeasures
are
secure.

\subsection{Our Contributions}
At
this
point
the
security
auditor
asks
``Can't
we
just
get
rid
of
speculative
execution?''---and
is
immediately
told
that
this
would
be
a
performance
disaster.
{\it
Every\/}
control-flow
instruction
would
cost
$P$
cycles
where
$P$
is
close
to
the
full
pipeline
length,
and
would
thus
cost
the
equivalent
of
$P\times
I$
instructions
where
$I$
is
the
number
of
instructions
per
cycle.
This
extra
$P\times
I$-instruction
cost
would
be
incurred
every
4--6
instructions.
The
emergency
security
patches
described
above
also
sacrificed
performance,
but
clearly
were
nowhere
near
{\it
this\/}
bad.

\relax

We
observe,
however,
that
this
performance
analysis
makes
an
implicit
assumption
regarding
the
instruction
set
architecture.
We
introduce
an
ISA
feature,
BasicBlocker,
that
undermines
this
assumption.
BasicBlocker
is
simple
and
can
be
efficiently
implemented
in
hardware.
We
show
how
modifications
to
the
compiler
utilize
the
BasicBlocker
design
to
minimize
the
performance
penalty
of
removing
not
only
branch
prediction,
but
also
speculative
fetching
(that
is,
instructions
are
fetched
but
never
executed)
from
a
processor.
The
resulting
processor
design
is
simpler
than
current
speculative
CPUs
which
removes
one
of
the
most
complicated
aspects
of
a
CPU
security
audit.

To
evaluate
performance
and
demonstrate
feasibility
of
Basic\-Blocker,
we
start
with
an
existing
compiler
and
an
existing
CPU
for
an
existing
ISA;
we
modify
all
of
these
to
support
BasicBlocker;
and
we
compare
the
performance
of
the
modified
CPU
to
the
performance
of
the
original
CPU.
We
selected
the
RISC-V
ISA~\cite{2014/asanovic}
given
its
openness.
To
demonstrate
the
compatibility
to
different
types
of
CPUs,
we
selected
two
implementation
platforms,
one
in-order
soft
core
(a
CPU
simulated
by
an
FPGA)
and
a
simulated
superscalar
out-of-order
processor
to
allow
evaluations
without
manufacturing
a
chip.
Full
details
of
our
BBRISC-V
ISA
appear
later
in
the
paper.

The
Spectre
authors
stated~\cite{kocher2019spectre}
that
they
``\textit{believe
that
long-term
solutions
will
require
fundamentally
changing
instruction
set
architectures}''.
Our
performance
results
rely
on
a
synergy
between
changes
to
the
CPU
and
changes
to
the
compiler,
mediated
by
changes
to
the
ISA\null.
To
improve
deployability,
we
explain
how
a
CPU
supporting
BasicBlocker
can
also
run
code
compiled
for
the
old
ISA.
Our
protection
against
Spectre
relies
solely
on
a
simple
change
to
the
CPU,
namely
disabling
speculation,
so
it
applies
both
to
old
code
and
to
new
code.
Recompilation
is
necessary
only
for
performance
reasons
to
relieve
occasional
hot
spots,
not
for
security.

\paragraph{Scope of This Work.}
Beyond
branch
prediction,
CPU
designers
have
added
many
forms
of
speculation
in
the
pursuit
of
every
last
bit
of
performance,
and
the
only
safe
assumption
is
that
every
form
of
speculation
threatens
security.
For
example,~\cite{koruyeh2018spectre}
exploits
the
prediction
of
return
addresses
and
\cite{horn2018speculative}
exploits
speculative
store-load
forwarding.

BasicBlocker
addresses
specifically
the
performance
loss
of
disabling
{\it
control-flow
speculation}.
This
includes
branch
prediction
and
return-address
speculation.
To
protect
against
attacks
exploiting
other
forms
of
speculation
(e.g.,
``Spectre-STL''),
we
recommend
that
the
CPU
designer
disable
all
forms
of
speculation,
not
just
control-flow
speculation.
This
is
easy
for
any
form
of
speculation
with
sufficiently
small
benefits,
but
otherwise
it
raises
ISA-design
challenges
and
performance-analysis
challenges.
Focusing
on
one
form
is
essential
to
make
the
analysis
tractable,
and
branch
prediction
in
particular
clearly
qualifies
as
an
important
target.

\subsection{The BasicBlocker Concept in a Nutshell}
The
$P$-cycle
branch-misprediction
cost
mentioned
above
is
the
time
from
early
in
the
pipeline,
when
instructions
are
fetched,
to
late
in
the
pipeline,
when
a
branch
instruction
computes
the
next
program
counter.
If
a
branch
passes
through
the
fetch
stage
and
is
mispredicted,
then
the
misprediction
will
not
be
known
until
$P$
cycles
later,
when
the
next
program
counter
is
computed.
Every
instruction
fetched
in
the
meantime
needs
to
be
rolled
back.

The
implicit
assumption
is
that
the
ISA
defines
the
branch
instruction
to
take
effect
starting
immediately
with
the
next
instruction.
This
assumption
was
already
challenged
by
``branch
delay
slots''
on
the
first
RISC
architecture
in
the
1980s;
see
generally~\cite{1987/derosa}.
A
branch
delay
slot
means
that
a
branch
takes
effect
only
after
the
next
instruction.
The
compiler
compensates
by
moving
the
branch
up
by
one
instruction,
if
there
is
an
independent
previous
instruction
in
the
basic
block,
the
contiguous
sequence
of
instructions
preceding
the
branch.
A
branch
delay
slot
reduces
the
cost
of
a
branch
misprediction
by
$1$
instruction,
and
the
first
RISC
CPU
pipeline
was
short
enough
that
this
removed
any
need
for
branch
prediction.

A
few
subsequent
CPUs
used
double
branch
delay
slots,
reducing
the
branch-misprediction
cost
by
$2$
instructions.
Obviously
one
can
define
an
architecture
with
$K=P\times
I$
delay
slots
after
each
branch.
However,
code
compiled
for
that
architecture
can
only
run
on
a
processor
with
exactly
$K$
delay
slots.
Since
an
optimal
$K$
depends
on
the
CPU,
code
would
have
to
be
compiled
for
every
target
CPU
individually.

In
a
BasicBlocker
ISA,
there
is
a
``basic
block
$N$''
instruction
guaranteeing
that
the
next
$N$
instructions\footnote
{It
is
natural
to
consider
a
variant
that
counts
$N$
fixed-length
words
(as
an
extreme,
$N$
bytes)
on
an
architecture
with
variable-length
instructions.}
will
all
be
executed
consecutively.
These
instructions
include,
optionally,
a
branch
instruction,
which
takes
effect
\textit{after}
the
$N$
instructions,
no
matter
where
the
branch
is
located
within
the
$N$
instructions.
The
same
ISA
supports
all
values
of
$N$
simultaneously.

It
is
the
CPU's
responsibility
to
disable
all
speculative
behavior,
including
speculative
fetching.
With
BasicBlocker,
most
of
the
performance
lost
from
disabling
control-flow
speculation
can
be
regained.
The
BasicBlocker
ISA
lets
the
compiler
declare
the
basic-block
size
and
move
the
branch
up
as
far
as
possible
within
the
block.
The
declaration
of
the
basic-block
size
lets
the
CPU
fetch
all
instructions
in
the
basic
block,
without
speculation.
If
the
branch
instruction
is
not
too
close
to
the
end
of
the
block
then
the
CPU
can
immediately
continue
with
the
next
basic
block,
again
without
speculation.
The
overall
performance
benefit
of
this
rescheduling
for
each
basic
block
matches
the
benefit
of
whatever
number
of
delay
slots
could
be
useful
for
that
microarchitecture,
without
the
disadvantage
of
having
to
be
compiled
differently
for
each
number
of
delay
slots.
The
new
instruction
further
allows
for
tight
integration
of
further
optimizations
such
as
hardware
loop
counters.

\section{Related Work}
\label{sec:related}

\subsubsection*{ISA Modifications.}
\label{designing-isas-for-security}
There
is
a
long
history
of
security
features
in
ISAs
including
extensions
to
enforce
control-flow
integrity
(CFI)~\cite{abadi2009control,
davi2015hafix},
memory
protection
(e.g.
ARM-MTE
\cite{armARF}),
or
the
flushing
of
microarchitectural
states~\cite{wistoff2020prevention}.
Other
extensions
simplify
the
secure
implementation
of
complicated
and
security-critical
aspects,
e.g.
by
adding
an
instruction
for
AES
computations~\cite{2010/gueron}.
All
these
ISA
extensions
introduce
new
instructions,
that
can
be
used
by
a
programmer
or
compiler
to
harden
a
program
against
some
specific
attacks.
Usage
of
the
new
features
(and
hence
the
protection)
requires
some
modification
of
the
binary
(mostly
through
recompilation),
but
unmodified
binaries
run
correctly
as
well.
In
all
cases
hardware
changes
are
required
to
support
the
new
instructions.

Some
ISAs
remove
incentives
for
control-flow
speculation,
although
not
motivated
by
security.
Berkley's
Precision
Timed
(PRET)
machines~\cite{Lee2017pret}
target
real-time
computing
applications
which
require
a
minimal
worst-case
runtime.
Hence,
control
flow
speculation
is
substituted
by
a
round-robin
scheduling
of
instructions
from
different
thread
contexts.
With
BasicBlocker
we
focus
on
single-threaded
applications
to
still
perform
well
without
control-flow
speculation,
but
thread
parallelism
is
likely
to
further
improve
performance.
VLIW
architectures~\cite{fisher1983very}
introduce
instruction
level
parallelism
by
explicitly
declaring
instructions
that
can
be
executed
in
parallel
at
compile
time.
VLIW
further
uses
compiler
heuristics
to
make
an
educated
guess
about
the
direction
of
a
branch.
If
the
branch
is
resolved
in
a
different
direction,
the
compiler
places
compensating
code
at
the
branch
target.
This
technique
relocates
the
speculation
problem
to
the
compiler
level.
A
major
drawback
of
VLIW
is
the
strict
compiler
dependency
on
the
target
platform:
many
microarchitecture
decisions
are
embedded
into
the
ISA,
and
code
must
be
recompiled
whenever
those
decisions
change.
BasicBlocker
is
carefully
designed
to
not
re-introduce
speculation
at
compiler
level
and
the
code
generated
by
the
compiler
does
not
depend
on
the
microarchitecture
of
the
target
CPU.

\subsubsection*{Spectre Countermeasures.}
Transient-execution
attacks,
including
speculative-execution
attacks,
gained
widespread
attention
after
the
disclosure
of
Spectre
\cite{kocher2019spectre}
and
Meltdown
\cite{lipp2018meltdown}.
The
attacks
in
\cite{kocher2019spectre, chen2019sgxpectre, koruyeh2018spectre, maisuradze2018ret2spec, lipp2018meltdown, schwarz2019zombieload, van2019ridl, canella2019fallout, van2018foreshadow, weisse2018foreshadow, van2019addendum}
have
shown
many
ways
that
transient
execution
can
undermine
memory
protection
and
violate
basic
security
assurances.
See~\cite{kiriansky2018speculative,szefer2019survey,Canella2019extended,canella2020evolution}
for
surveys
of
attack
vectors
and
countermeasures.
In
the
following
we
will
focus
on
countermeasures
against
control-flow
speculation
based
attacks.
Typically,
such
attacks
arrange
for
mispredicted
instructions
to
access
sensitive
data.
The
instructions
are
eventually
rolled
back
but
still
leave
footprints
in
the
microarchitectural
state.

The
countermeasures
presented
in
\cite{turner2018retpoline,
zhao2020lightweight}
prevent
the
attacker
from
controlling
the
branch
prediction.
Such
countermeasures
are
specialized
to
prevent
a
specific
type
of
Spectre
attack
in
a
specific
setting.
Other
approaches
close
a
specific
covert
channel,
most
prominently
the
timing
channel
introduced
through
caches~\cite{yan2018invisispec,
Khasawneh.2018,
kiriansky2018dawg,
li2019conditional,
sakalis2019efficient,
aciiccmez2010new,
braun2015robust,zhang2013duppel,
varadarajan2014scheduler,
wistoff2020prevention}.
Again
those
countermeasures
are
targeted
at
a
specific
setting
and
other
covert
channels
remain
exploitable.

A
more
general
approach
of
countermeasures
targets
the
attackers
ability
to
create
a
secret-dependent,
transient
CPU
state
in
combination
with
a
covert
channel.
This
can
be
done
by
limiting
the
microarchitectural
operations
that
can
be
performed
on
sensitive
values~\cite{yu2019data,zagieboylo2019using,schwarz2019context,
barber2019specshield,
Yu.2019,
weisse2019nda}.
Such
approaches
require
the
knowledge
which
values
are
considered
as
secret
as
well
as
a
model
that
defines
which
kind
of
behavior
(instructions
or
group
of
instructions
in
a
transient
setting)
is
dangerous.
The
security
and
performance
overhead
is
highly
dependent
on
the
selection
of
this
security
model
and
the
definition
is
not
trivial,
as
new
channels
are
discovered
constantly
(see,
e.g.,
\cite{Behnia.2020}).
Reported
overheads
reach
from
$10\%$
\cite{barber2019specshield}
to
$125\%$
\cite{weisse2019nda},
but
require
the
consideration
of
the
specific
measurement
environment.

Like
most
of
the
cited
countermeasures,
BasicBlocker
requires
changes
to
the
hardware
mediated
by
the
ISA.
In
contrast
to
other
approaches,
BasicBlocker
does
not
aim
to
fix
the
problems
induced
by
control-flow
speculation,
but
rather
tries
to
mitigate
the
performance
penalty
caused
by
removing
control-flow
speculation
entirely.
The
reasoning
behind
this
approach
is
that
only
the
removal
of
speculative
behavior
is
guaranteed
to
remove
all
speculation-based
attack
vectors,
by
removing
the
root
cause
of
the
vulnerability.
The
comparability
of
the
resulting
performance
overhead
is
limited,
as
we
also
consider
the
impact
of
speculative
fetching,
which
is
mostly
ignored
by
state-of-the-art
Spectre
countermeasures.

This
paper
focuses
on
speculative-execution
attacks.
It
should
be
possible
to
similarly
address
fault-based,
transient-execution
attacks
by
``preponing''
fault
detection,
removing
most
of
the
performance
benefit
of
transient
execution
after
faults,
but
further
investigation
of
this
idea
is
left
to
future
work.

\relax

\section{Speculation in Processors}
\label{sec:spec}

In
a
pipelined
processor,
each
instruction
passes
through
multiple
pipeline
stages
before
it
eventually
retires.
A
textbook
series
of
stages
is
\emph{Instruction Fetch} (IF), \emph{Instruction Decode} (ID), \emph{Execution} (EX),
\emph{Memory Access} (MEM) and \emph{Write Back} (WB) \cite{tanenbaum2016structured}.
More
complex
CPUs
can
have
many
more
stages.

If
each
stage
takes
one
cycle
then
a
branch
instruction
will
be
fetched
on
cycle
$n$
in
IF,
decoded
on
cycle
$n+1$
in
ID,
and
executed
on
cycle
$n+2$
in
EX,
so
at
the
end
of
cycle
$n+2$
the
CPU
knows
whether
the
branch
is
taken
or
not.
Without
branch
prediction,
IF
stalls
on
cycles
$n+1$
and
$n+2$,
because
it
does
not
know
yet
which
instructions
to
fetch
after
the
branch.
With
branch
prediction,
IF
speculatively
fetches
instructions
on
cycles
$n+1$
and
$n+2$,
and
ID
speculatively
decodes
the
first
of
those
instructions
on
cycle
$n+2$.
If
the
prediction
turns
out
to
be
wrong
then
the
speculatively
executed
instructions
are
\textit{rolled
back\/}:
all
of
their
intermediate
results
are
removed
from
the
pipeline.

The
functional
effects
of
instructions
are
visible
only
when
the
instructions
retire,
but
side
channels
sometimes
reveal
microarchitectural
effects
of
instructions
that
have
been
rolled
back.
As
Spectre
illustrates,
this
complicates
the
security
analysis:
one
can
no
longer
trust
a
branch
to
stop
the
wrong
instructions
from
being
visibly
partially
executed.

\begin{figure*}[t]
\begin{minipage}{.48\textwidth}
\begin{lstlisting}[language=riscv]
; Start of first basic block
add	a5,a0,a4
add	t4,a3,a4
addi	a4,a4,8
mul	a1,t3,t2
lh	t2,0(a5)
bne	a4,a6,80... ; compute branch and change PC
; Start of 2nd basic block
lh	a7,0(a1)
li	a4,0
; Start of 3rd basic block
sh	a1,0(a0)
...	
\end{lstlisting}
\end{minipage} \quad
\begin{minipage}{.48\textwidth}
\begin{lstlisting}[language=riscv]
bb	6, 0 ; first bb, size = 6, not seq
add	a5,a0,a4
add	t4,a3,a4
addi	a4,a4,8
bne	a4,a6,80... ; compute branch result
mul	a1,t3,t2
lh	t2,0(a5) ; change PC after this instr.
bb	2, 1 ; 2nd bb, size = 2, seq
lh	a7,0(a1)
li	a4,0
bb	16, 0 ; 3rd bb, size = 16, not seq
sh	a1,0(a0)
...
\end{lstlisting}
\end{minipage}
\caption{\label{mini:bb_instr}Example code for the new \texttt{bb} instruction. Left: Traditional RISC-V code does not contain information about the size of upcoming basic blocks. The \texttt{bne} instruction terminates the first block and conditionally branches. Right: The \texttt{bb} instruction gives information about upcoming code parts. The first basic block is terminated by the size given in the line 1 and performs a conditional branch based on the outcome of the \texttt{bne} instruction, whose result is already determined earlier.}
\end{figure*}

The
standard
separation
of
fetch
from
decode
also
means
that
{\it
every
instruction
is
being
speculatively
fetched}.
An
instruction
fetched
in
cycle
$n$
could
be
a
branch
(or
other
control-flow
instruction),
but
the
CPU
knows
this
only
after
ID
decodes
the
instruction
in
cycle
$n+1$,
so
IF
is
speculatively
fetching
an
instruction
in
cycle
$n+1$.
We
emphasize
that
this
behavior
is
present
even
on
CPUs
without
branch
prediction:
the
CPU
cannot
know
whether
the
instruction
changes
the
control
flow
before
decoding
it.

Disabling
all
control-flow
speculation
execution
thus
means
that
every
branch
must
stall
fetching
until
it
is
executed,
and,
perhaps
even
more
importantly,
that
{\it
every
instruction\/}
must
stall
fetching
until
it
is
decoded.
BasicBlocker
addresses
both
of
these
performance
problems,
as
shown
below.

\section{Concept}
\label{sec:concept}
In
this
section,
we
outline
the
rationale
behind
our
approach
as
well
as
the
modifications
to
the
ISA
that
allow
the
elimination
of
control-flow
speculation
within
the
microarchitecture.
Though
we
use
the
RISC-V
instruction
set
in
the
following
examples,
our
solution
is
generally
applicable
to
any
ISA
or
processor
as
motivated
in
Section~\ref{sec:compat}
and
\ref{sec:generic}.

\subsection{Design Rationale}
It
is
conceptually
simple
to
generically
thwart
security
issues
arising
from
control-flow
speculation
by
entirely
removing
it,
but
is
generally
believed
to
incur
a
severe
loss
in
performance.
BasicBlocker
addresses
this
by
providing
metadata
through
an
ISA
modification
to
assist
non-speculative
hardware
with
efficient
execution
of
software
programs.

The
CPU
has
a
limited
view
of
programs,
accessing
only
a
limited
number
of
instructions
at
a
time.
With
current
ISAs,
control-flow
instructions
appear
without
advance
notice,
and
their
result
is
available
only
after
multiple
pipeline
stages,
even
though
this
result
is
needed
immediately
to
infer
the
next
instruction.

BasicBlocker
takes
the
concept
of
basic
blocks
(in
contrast
to
the
textbook
definition,
we
require
a
basic
block
to
be
terminated
by
all
control-flow
instructions,
i.e.
also
calls)
to
the
hardware
level
using
novel
instructions.
At
compile
time
a
holistic
view
of
the
program
is
available
in
form
of
a
control-flow
graph,
including
code
structure
such
as
basic
blocks
and
control-flow
changes.
BasicBlocker
uses
the
information
available
at
compile
time,
specifically
the
length
of
individual
basic
blocks,
and
makes
it
available
to
the
CPU
during
execution.
This
allows
a
non-speculative
CPU
to
avoid
most
pipeline
stalls,
through
the
advance
notice
of
control
flow
changes.

\subsection{Basic Block Instruction}
\label{sec:basicBlockInstruction}
We
introduce
a
new
instruction,
called
basic
block
instruction
(\texttt{bb}),
which
lays
the
foundation
for
BasicBlocker.
Currently,
most
CPUs
use
control-flow
speculation
to
gain
performance.
Enabling
fast
but
non-speculative
fetching
requires
additional
information
for
the
CPU,
since
normally
we
know
that
we
can
fetch
the
next
instruction
only
after
the
prior
instruction
was
decoded
and
it
is
ensured
that
the
control
flow
does
not
deviate.
Hence,
normally
the
fetch
unit
would
have
to
be
stalled
until
the
previous
instruction
was
decoded.
To
avoid
that
delay,
we
define
a
new
\texttt{bb}
instruction
that
encodes
the
size
of
the
basic
block.
Within
this
basic
block,
the
CPU
is
allowed
to
fetch
instructions,
knowing
that
upcoming
instructions
can
be
found
in
sequential
order
in
memory
and
will
definitely
be
executed.
That
is,
since
per
definition
no
control
flow
changes
can
occur
within
the
basic
block.
The
instruction
further
provides
information
whether
the
basic
block
is
\textit{sequential},
stating
that
the
control
flow
continues
with
the
next
basic
block
in
the
sequence
in
memory,
i.e.
the
block
does
not
contain
a
control-flow
instruction.
Figure
\ref{mini:bb_instr}
shows
the
transformation
of
traditional
code
(left)
to
code
with
\texttt{bb}
instructions
(right).
The
fetch
unit
of
the
CPU
is
responsible
for
counting
the
remaining
instructions
in
a
given
block
and
only
fetch
until
the
end
of
the
basic
block.
From
there,
the
program
continues
executing
the
next
basic
block
which
itself
starts
with
a
\texttt{bb}
instruction.

We
also
modified
the
behavior
of
existing
control-flow
instructions,
such
as
\texttt{bne},
\texttt{j}
and
\texttt{jlre}.
The
goal
is
to
give
advance
notice
of
upcoming
control-flow
changes
to
the
CPU.
Since
the
processor
knows
the
number
of
remaining
instructions
per
basic
block,
we
can
schedule
control-flow
instructions
within
basic
blocks
as
early
as
data
dependencies
allow,
and
still
perform
the
change
of
the
control
flow
at
the
end
of
the
basic
block.
This
key
feature
allows
the
CPU
to
correctly
determine
the
control
flow
before
the
end
of
the
basic
block,
and
renders
branch
prediction
in
many
cases
obsolete.

As
a
result,
the
only
time
that
the
CPU
needs
to
stall
fetching
is
at
the
transition
of
two
basic
blocks,
because
the
following
\texttt{bb}
instruction
needs
to
be
executed
before
knowing
the
size
and,
hence,
being
able
to
continue
fetching.
To
avoid
this
delay,
it
is
sufficient
to
add
the
capability
of
representing
one
additional
set
of
basic
block
information
internally
and
request
this
information
as
early
as
possible.
This
means
that
the
CPU
interposes
the
\texttt{bb}
instruction
of
the
next
basic
block
as
soon
as
the
next
basic
block
is
known,
regardless
whether
there
are
instructions
left
in
the
current
basic
block
or
not.

\begin{figure} [h]
\centering
\includestandalone[width=.9\linewidth]{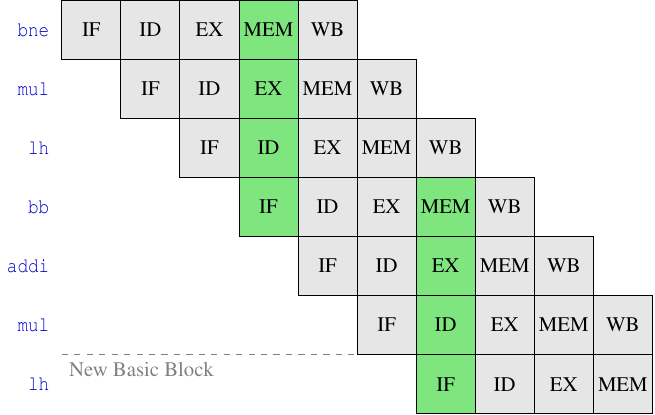}
\caption{\label{tikz:pipeline_bb_preexecute}Pipeline diagram for optimal code. The \texttt{bb} instruction of the next basic block is fetched as soon as the branch was executed. The branch only takes effect at the end of the current basic block. When the branch instruction is sufficiently early rescheduled, the next basic block can be fetched without stalls.}
\end{figure}

In
Figure
\ref{tikz:pipeline_bb_preexecute},
this
principle
is
illustrated
for
the
code
of
Figure
\ref{mini:bb_instr}
(right
side).
The
\texttt{bb}
instruction
of
the
second
basic
block
is
fetched
as
soon
as
the
branch
target
of
\texttt{bne}
is
known.
Afterwards,
the
execution
of
the
first
basic
block
continues.
Execution
of
the
second
basic
block
can
start
as
soon
as
the
first
basic
block
is
consumed
and
the
size
of
the
second
basic
block
is
known
(after
EX
of
\texttt{bb}).
If
the
current
basic
block
does
not
contain
a
control-flow
instruction,
which
is
indicated
by
the
\textit{sequential}
flag
of
the
\texttt{bb}
instruction,
the
CPU
can
fetch
the
next
\texttt{bb}
instruction
directly.
Otherwise,
the
next
\texttt{bb}
instruction
will
be
fetched
after
the
control-flow
instruction
passes
the
execution
stage.

While
the
early
fetching
of
the
\texttt{bb}
instruction
changes
the
execution
order,
it
does
not
affect
security
or
correctness
since
the
instruction
is
only
fetched
after
the
execution
path
is
known
for
certain.

\begin{figure} [h]
\centering
\includestandalone[width=.9\linewidth]{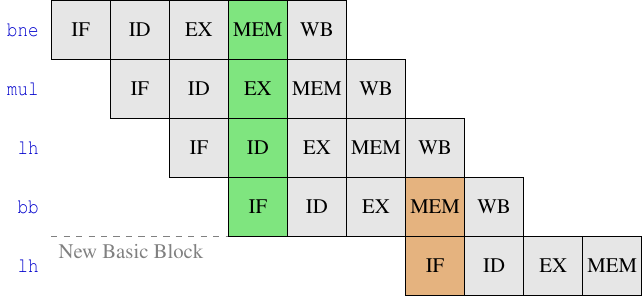}
\caption{\label{tikz:pipeline_concept}Pipeline diagram for code with non-optimal rescheduling of branch instructions. The next \texttt{bb} instruction is not finished with execution when the new basic block begins. The CPU needs to stall until the basic block size is known which is generally after the execution stage.}
\end{figure}

Even
with
these
changes
it
is
necessary
to
stall
the
CPU
at
the
transition
of
two
basic
blocks
until
the
size
of
the
new
basic
block
is
known.
Therefore,
this
concept
works
best
with
software
that
contains
many
large
basic
blocks
with
multiple
opportunities
to
\textit{reschedule} control-flow instructions at compile time. Software with a large number
of
small
basic
blocks
is
therefore
less
efficient,
leading
to
pipeline
stalls
as
shown
in
Figure
\ref{tikz:pipeline_concept}.

\begin{figure} [H]
\centering
\includestandalone[width=.9\linewidth]{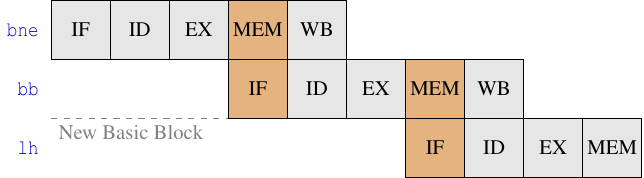}
\caption{\label{tikz:pipeline_worst_case}The worst case scenario has a branch instruction at the end of a basic block.}
\end{figure}

The
worst
case
is
a
control-flow
instruction
that
could
not
be
rescheduled,
since
then
the
CPU
needs
to
be
stalled
both
for
the
information
from
the
control-flow
instruction
as
well
as
from
the
\texttt{bb}
instructions.
This
case
is
depicted
in
Figure
\ref{tikz:pipeline_worst_case}.
We
address
the
performance
impact
of
small
basic
blocks
in
Section
\ref{sec:concept_optimization}.

Overall,
the
rescheduling
concept
can
be
imagined
as
a
variably-sized
branch
delay
slot.
There
are
two
core
advantages
of
our
concept
over
traditional
branch
delay
slots:
\begin{itemize}
\item The CPU does not need special constructs for the branch delay instructions. At the end of a basic block, the CPU can simply fetch the instruction at the target address, regardless of the type of instructions that were executed prior. If the basic block was sequential, the target register defaults to $PC+4$. If any control-flow operations were executed, the target register points to the target address.
\item By having a variably-sized branch delay mechanism, the code is compatible to all hardware architectures that support the \texttt{bb} instruction. Since the control-flow instructions were rescheduled as early as possible, the code is optimal for those hardware architectures. For fixed size branch delay slots, CPUs with smaller pipelines may introduce unnecessary \texttt{nop} instructions.
\end{itemize}
See
also
Section~\ref{sec:concept_optimization}
for
further
optimizations
that
integrate
tightly
with
the~\texttt{bb}
instruction.

\subsubsection{ISA-Extension Specification}
We
now
define
the
changes
required
by
BasicBlocker
more
precisely.
A
processor
supporting
the
\texttt{bb}
instruction
is
required
to
have
an
instruction
counter
$IC$,
a
target
register
$T$,
a
branch
flag
$B$,
and
an
exception
flag
$E$,
all
initialized
to
$0$
on
processor
reset
and
used
only
as
defined
below.
The
functional
behavior
of
the
\texttt{bb}
instruction
is
given
in
Definition
\ref{def:bb_f},
the
changes
to
the
control
flow
in
Definition
\ref{def:bb_s}
and
the
behavior
that
raises
an
exception
in
Definition
\ref{def:bb_exception}.

\begin{definition}[BB Instruction]
\label{def:bb_f}
The
\texttt{bb}
instruction
takes
a
size
parameter
$n>0$
and
a
sequential
flag
$seq$,
and
is
executed
as
follows.
If
$IC=0$:
$IC\gets
n$;
if
$seq=0$
then
$B\gets
1$;
if
$seq=1$
then
$B\gets
0$
and
$T$
is
set
to
the
address
of
the
$n+1$-th
instruction
following
the
\texttt{bb}
instruction.
Otherwise,
if
$IC\ne
0$,
then
$IC\gets
0$
and
$E\gets
1$
to
catch
illegal
\texttt{bb}
instructions.
\end{definition}

Thus,
on
a
functional
level,
Definition~\ref{def:bb_f}
only
sets
$IC$,
$T$,
$B$,
and
$E$
but
has
no
further
effect
on
the
execution
of
a
program.
The
subsequent
definitions
have
further
effects.

\begin{definition}[BB-Delayed Branches]
\label{def:bb_s}
The
execution
of
non-\texttt{bb}
instructions
is
modified
as
follows:
\begin{itemize}
\item
Before
every
non-\texttt{bb}
instruction:
if
$IC>0$
then
$IC\gets
IC-1$.
\item
During
every
control-flow
instruction:
any
write
to
$PC$
is
instead
written
to
$T$
if
$B>0$,
and
is
ignored
if
$B=0$.
\item
After
every
control-flow
instruction:
if
$B=0$
then
$E\gets
1$;
otherwise
$B\gets
B-1$.
\item
Subsequently,
after
every
non-\texttt{bb}
instruction:
if
$IC=0$
then
$PC\gets
T$;
and
if
$IC=0$
and
$B>0$
then
$E\gets
1$.
\end{itemize}
\end{definition}

BasicBlocker
raises
an
exception
($E=1$)
whenever
the
\texttt{bb}
instruction
is
used
in
an
illegal
way.

\begin{definition}[BB Exceptions]
\label{def:bb_exception}
After
every
instruction,
an
exception
is
raised
if
$IC=0$
and
$E\ne
0$.
\end{definition}

In
other
words,
after
the
$n$
instructions
covered
by
a
\texttt{bb}
instruction,
an
exception
is
raised
if
any
of
the
following
occurred:
\begin{itemize}
	\item
$seq=0$
and
there
was
not
exactly
one
control-flow
instruction
in
the
$n$
instructions;
	\item
$seq=1$
and
there
was
a
control-flow
instruction
in
the
$n$
instructions.
	\item
A
\texttt{bb}
instruction
appears
within
the
$n$
instructions
indicated
by
the
previous
\texttt{bb}
instruction.
\end{itemize}

All
three
definitions
are
required,
in
order
to
add
BasicBlocker
to
an
arbitrary
ISA.
The
following
extra
requirement,
a
requirement
to
use
\texttt{bb}
instructions,
slightly
simplifies
the
implementation
of
BasicBlocker,
although
later
we
consider
dropping
this
requirement
for
compatibility.

\begin{definition}[Enforced BB]
\label{def:bb_required}
In
a
BasicBlocker
CPU
with
enforced
BB:
Before
every
non-\texttt{bb}
instruction
(and
before
$IC$
is
decremented),
an
exception
is
raised
if
$IC=0$.
\end{definition}

To
achieve
an
increased
performance,
an
implementation
of
BasicBlocker
can
pre-execute
\texttt{bb} instructions (cf. Figure \ref{tikz:pipeline_bb_preexecute}) as defined in
Definition
\ref{def:prefetch}.
This
pre-execution
affects
the
microarchitecture
and
timing
but
not
the
ISA
semantics.

\begin{definition}[BB Prefetching]
\label{def:prefetch}
A
BasicBlocker
CPU
with
prefetching
pre-executes
a
\texttt{bb}
instruction
$bb_{i+1}$
during
the
execution
of
a
block,
indicated
by
the
\texttt{bb}
instruction
$bb_i$,
as
soon
as:
\begin{itemize}
\item if $bb_i$ is sequential: $bb_i$ is resolved.
\item if $bb_i$ is not sequential: the first control flow instruction of the block is resolved.
\end{itemize}
This
requires
an
additional
register
$P$
which
holds
the
values
$n$
and
$seq$
until
execution
reaches
the
instruction
following
the
prefetched
\texttt{bb}
instruction.
More
precisely,
when
$IC=0$
and
$E=0$:
\begin{itemize}
\item $IC \gets n$ taken from $P$.
\item if $seq=0$ in P than $B \gets 1$ else $B \gets 0$.
\end{itemize}
If
the
prefetch
address
is
invalid,
or
if
the
prefetch
address
is
valid
but
the
prefetched
instruction
is
not
a
\texttt{bb}
instruction,
then
pre-execution
is
skipped
and
does
not
raise
an
exception.
\end{definition}

\subsection{Further Optimizations}
\label{sec:concept_optimization}
The
above
presented
concept
can
be
further
optimized
by
providing
the
information
contained
in
the
\texttt{bb}
instruction
as
soon
as
possible
using
pipeline
forwarding.
By
construction,
none
of
the
information
contained
in
the
\texttt{bb}
instructions
affects
any
other
element
of
the
CPU
than
the
fetch
unit.
Hence,
it
is
possible
to
wire
these
bits
back
to
the
fetch
unit
directly
after
the
decode
stage
without
further
changes
to
the
design.
Another
clock
cycle
can
be
saved
by
using
a
bit
mask
to
fast-decode
the
output
of
the
instruction
memory
directly,
with
only
marginal
overhead.

A
significant
boost
for
performance
can
be
achieved
by
introducing
an
additional
interface
to
the
instruction
memory
(or
cache)
that
is
used
to
access
\texttt{bb}
instructions.
This
would
allow
the
fetch
unit
to
request
and
process
\texttt{bb}
instructions
in
parallel
with
the
normal
instructions
and,
therefore,
eliminate
the
entire
performance
overhead
that
is
introduced
though
the
addition
of
these
instructions.
Since
a
basic
block
contains
always
at
least
one
instruction
additional
to
the
\texttt{bb}
instruction,
this
instruction
can
be
fetched
before
knowing
the
size
of
the
basic
block,
without
violating
the
above
stated
principles.

Further
optimizations
are
possible
with
additional
changes
to
the
ISA.
For
example,
the
1-bit
sequential
flag
can
be
replaced
by
a
multi-bit
counter
of
the
number
of
control-flow
instructions
in
the
upcoming
block,
so
(e.g.)~{\catcode`\&=12\tt
if(a&&b&&c)}
can
be
expressed
as
three
branches
out
of
a
single
block.
This
also
changes
the
branch
flag
$B$
to
a
multi-bit
branch
counter.

The
idea
to
announce
upcoming
control-flow
changes
early
on
is
also
the
foundation
of
hardware
loop
counters,
as
already
discussed
in
the
literature
\cite{dipasquale2003hardware, raghavan2008distributed}. Here, the software announces
a
loop
to
the
hardware,
which
then
takes
responsibility
for
the
correct
execution.
We
can
seamlessly
support
hardware
loop
counters
in
our
design
concept.
One
new
instruction
(\texttt{lcnt})
is
necessary
to
store
the
number
of
loop
iterations
into
a
dedicated
register.
The
start
and
end
address
of
a
loop
can
be
encoded
into
the
\texttt{bb}
instruction,
by
indicating
with
two
separate
flags
whether
the
corresponding
basic
block
is
the
start
(s-flag)
or
end
(e-flag)
block
of
the
loop.
This
allows
the
hardware
to
know
the
next
basic
block,
as
soon
as
the
\texttt{bb}
instruction
of
the
end
block
gets
executed.
The
fast
execution
of
nested
loops
can
be
supported
by
adding
multiple
start
and
end
flags
to
the
\texttt{bb}
instruction
as
well
as
adding
multiple
registers
for
the
number
of
loop
iterations.
A
more
detailed
description
of
the
loop
counter
integration
to
our
concept
can
be
found
in
Appendix
\ref{appendix:loopCounter}.

\subsection{Compatibility}
\label{sec:compat}

For
simplicity
and
comprehension
all
examples
above
consider
an
in-order,
single
issue
processor
with
a
generic
five
stage
RISC
pipeline.
Control-flow
speculation
is
widely
used
in
such
processors:
e.g.,
the
ARM
Cortex-A53,
which
has
shown
to
be
vulnerable
against
speculative-execution
attacks
\cite{nemati2020speculative}.
There
is
also
tremendous
interest
in
larger,
super-scalar,
out-of-order
processors,
where
control-flow
speculation
is
universal.

Adding
support
for
out-of-order
processors
is
trivial
as
per
design,
every
instruction
that
is
fetched
by
the
processor
will
be
retired
-
that
is,
if
none
of
the
instructions
raise
an
exception.
Once
the
CPU
fetches
the
instruction,
reordering
is
permitted
as
far
as
functional
correctness
is
ensured.
Utilizing
the
two
counter
sets,
reordering
can
be
done
beyond
basic
block
borders
if
the
\texttt{bb}
instruction
of
the
following
basic
block
has
been
executed.

Similarly,
support
for
superscalarity
is
easy
to
achieve.
Once
the
\texttt{bb}
instruction
is
executed,
the
CPU
may
fetch
and
execute
all
instructions
within
the
current
basic
block
in
an
arbitrary
amount
of
cycles.
If
the
successor
basic
block
is
known
the
CPU
may
fetch
instructions
from
both
basic
blocks
in
one
cycle.
Secondary
pipelines
may
also
be
useful
to
pre-execute
\texttt{bb}
instructions
for
the
following
basic
block
in
parallel
as
described
earlier.

Generally,
the
pipeline
length
can
be
chosen
flexibly.
However,
as
the
CPU
needs
to
wait
for
results
of
branch
and
\texttt{bb}
instructions,
it
is
desirable
to
make
the
results
of
these
instructions
available
as
early
as
possible.

A
major
feature
of
modern
systems
is
the
support
of
interrupts
and
context
switches.
We
note
that
our
concept
does
not
impede
such
features;
it
merely
increases
the
necessary
CPU
state
that
needs
to
be
saved
in
such
an
event.
More
specifically,
it
is
necessary
to
save
the
already
gathered
information
about
the
current
and
upcoming
basic
blocks
as
well
as
the
state
of
the
loop
counter,
in
addition
to
all
information
usually
saved
during
a
context
switch.
It
is
important
that
this
data
is
secured
against
manipulation
but
that
is
true
for
all
data
stored
during
a
context
switch
(e.g.
register
values,
FPU
state,
...).

Our
proposal
includes
one
new
instruction
and
a
modification
to
existing
control-flow
instructions.
For
easier
deployability,
it
is
desirable
for
a
BasicBlocker
CPU
to
be
backwards-compatible.
One
could
define
new
BasicBlocker
control-flow
instructions
separate
from
the
previous
control-flow
instructions.
However,
it
suffices
to
interpret
a
control-flow
instruction
as
having
the
new
semantics
if
it
is
within
the
range
of
a
\texttt{bb}
instruction,
and
otherwise
as
having
the
old
semantics,
dropping
Definition~\ref{def:bb_required}.
Legacy
code
compiled
for
the
non-BasicBlocker
ISA
will
then
run
correctly
but
with
low
performance,
and
code
recompiled
to
use
\texttt{bb}
will
run
correctly
with
high
performance.

It
would
also
be
possible
to
integrate
our
solution
into
a
secure
enclave
by
providing
a
modified
fetch
unit
for
the
enclave.
Security
critical
applications
could
be
run
in
the
protected
enclave
while
legacy
software
can
be
executed
on
the
main
processor
without
performance
losses.

\subsection{BasicBlocker for Generic ISAs}
\label{sec:generic}

In
the
following,
we
outline
the
changes
necessary
to
implement
the
BasicBlocker
concept
in
arbitrary
ISAs.
We
observe
that
in
common
ISAs,
branches
are
realized
with
three
basic
operations
which
are
performed
by
a
varying
number
of
instructions.

\begin{enumerate}
\item 	First, the operands on which the branch decision will be made are
	compared.
The
result
of
the
comparison
may
be
saved
in
a
	special
purpose
flag
(e.g.
Intel
x86,
ARM),
a
register
value,
or
	used
immediately
(e.g.
RISC-V,
some
Intel
x86).
\item	Based on the outcome of the comparison, the target address is
	computed.
\item 	The instruction pointer is changed to the target address computed
	in
the
previous
stage.
\end{enumerate}

For
most
ISAs,
steps
2)
and
3)
are
combined
to
one
instruction.
RISC-V
is
unusual
in
having
\textit{only}
branch
instructions
that
combine
all
three
operations.
	
A
BasicBlocker
ISA
is
required
to
separate
operation
1)
and
2)
from
3),
thus
avoiding
the
need
for
speculative
instruction
fetching.
Hence,
a
BasicBlocker
ISA
needs
at
least
one
instruction
that
compares
the
operands
and
computes
the
target
address.
Operation
3)
is
handled
implicitly
by
the
\texttt{bb}
instruction
at
the
beginning
of
the
basic
block,
which
indicates
after
how
many
instructions
the
instruction
pointer
is
updated
to
the
target
register.
A
BasicBlocker
ISA
\textit{may}
separate
operation
1)
and
2)
arbitrarily.
For
example,
an
ARM
version
of
BasicBlocker
could
keep
the
decoupled
compare
instruction.
The
branch
instructions
would
only
compute
the
target
address
based
on
the
compare
and
the
instruction
pointer
would
be
updated
to
the
target
address
at
the
end
of
a
basic
block,
indicated
by
the
previous
\texttt{bb}
instruction.

\subsection{Security}
BasicBlocker
was
carefully
designed
with
security
in
mind
and
the
following
section
provides
an
overview
of
the
security
argument.

\subsubsection{Defense Against Spectre-type Attacks}
The
first
and
foremost
goal
of
BasicBlocker
is
to
allow
removing
control-flow
speculation
to
prevent
Spectre-type
attacks.
CPUs
that
implement
BasicBlocker
should
be
designed
after
the
following
principle:
\begin{myindentpar}{.5cm}
\textbf{\textit{The microarchitectural state of a CPU is
affected
only
by
instructions
that
will
eventually
be
retired.}}
\end{myindentpar}
Processors
adhering
to
this
principle
are
not
allowed
to
do
any
type
of
control-flow
speculation,
including
speculative
fetching,
as
speculation
always
affects
the
microarchitectural
state
at
least
temporarily.
This
strict
and
simple
design
principle
leads
directly
to
the
conclusion
that
the
CPU
is
not
vulnerable
against
any
Spectre-type
attack
exploiting
control-flow
speculation,
including
\emph{Spectre-PHT},
\emph{Spectre-BPB},
and
\emph{Spectre-RSB}
(taking
the
classification
of
\cite{Canella2019extended}).
BasicBlocker
enables
fast
and
efficient
execution
of
code
while
maintaining
the
above
stated
principle.

Since
BasicBlocker
inherently
does
not
provide
mechanisms
targeting
the
performance
impact
of
disabling
data-flow
speculation
(e.g.
store-load
forwarding,
data
cache
prefetching),
we
consider
attacks
exploiting
data-flow
speculation
such
as
\emph{Spectre-STL}
(again
taking
the
classification
of
\cite{Canella2019extended})
out
of
scope
for
this
paper.
It
is
the
CPU
designer's
responsibility
to
prevent
exploitation
of
data-flow
speculation
which
can
either
be
achieved
by
disabling
it
entirely
or
by
implementing
appropriate
countermeasures.
It
is
also
possible
to
extend
BasicBlocker
to
provide
performance
recovering
mechanisms
for
data-flow
speculation,
e.g.
by
flagging
allowed
store
to
load
forwarding
code
constructs
at
compile
time,
but
we
leave
this
for
future
work.
We
also
do
not
discuss
exception-based
attacks
such
as
Meltdown~\cite{lipp2018meltdown}.

\subsubsection{Manipulation of BB Instruction Arguments}
In
the
following,
we
consider
a
powerful
attacker
that
is
able
to
manipulate
the
\texttt{bb} instruction arguments or the internal state of the \texttt{bb} registers.
An
attacker
able
to
manipulate
arguments
of
the
\texttt{bb}
instruction
is
in
control
of
certain
parts
of
the
control
flow,
by
either
flipping
the
sequential
flag,
decreasing
the
basic
block
size,
or
increasing
the
basic
block
size.
Flipping
the
sequential
flag
will
always
lead
to
an
exception,
due
to
Definition
\ref{def:bb_exception}.
Decreasing
the
basic
block
size
allows
to
skip
the
last
instructions
of
a
basic
block,
which
might
be
critical,
e.g.
the
removal
of
a
secret
key.
Increasing
the
basic
block
size
raises
an
exception
in
the
\emph{enforced
BB}
mode
(Definition
\ref{def:bb_required}),
but
allows
the
execution
of
additional
instructions
in
the
legacy
mode.
Such
additional
instructions
might
be
sufficient
to
form
a
covert
channel,
if
the
required
gadgets
can
be
found
in
the
executable.

While
those
attacks
may
be
harmful,
this
attacker
model
requires
full
control
over
the
code
executed
on
the
victim's
device
and/or
the
register
state.
Generally,
there
are
two
points
in
time
where
an
attacker
can
inject
the
manipulations
described
above:
1)
at
compile
time
and
2)
at
runtime.
For
1),
the
attacker
must
be
in
control
of
the
compiler
which
gives
full
control
over
the
code
anyway.
In
addition,
a
simple
static
analysis
is
sufficient
to
verify
the
correctness
of
all
\texttt{bb}
arguments
of
a
specific
binary.
2)
Manipulation
at
run
time
comes
down
to
either
code
injection
or
manipulation
of
internal
values
of
the
CPU
for
a
particular
program
state,
e.g.
during
a
context
switch
or
physical
fault
attack.
Both,
an
attacker
in
control
of
the
register
state
and
an
attacker
able
to
perform
code
injection,
have
full
control
over
the
code
executed
by
the
victim's
device
in
any
case.
BasicBlocker
does
not
affect
important
OS
security
features
like
access
rights
management
and
therefore
does
not
facilitate
such
attacks.

\section{Implementation}
\label{sec:impl}

We
now
give
a
specific
example
of
BasicBlocker
applied
to
an
ISA,
by
defining
BBRISC-V,
a
BasicBlocker
modification
of
the
RISC-V
ISA.
We
further
present
a
proof-of-concept
implementation
on
a
BBRISC-V
soft
core
as
well
as
a
timing
accurate
simulator.
To
allow
running
a
variety
of
benchmarks,
we
also
provide
a
modified
compiler
for
the
BBRISC-V
ISA.

Our
modified
ISA
additionally
specifies
support
for
hardware
loop
counters,
as
proposed
in
Section
\ref{sec:concept_optimization},
which
we
partly
evaluate
in
Appendix
\ref{appendix:loopCounter}.

\subsection{BBRISC-V ISA}

\begin{figure*} [h]
\centering
\includestandalone[width=.8\textwidth]{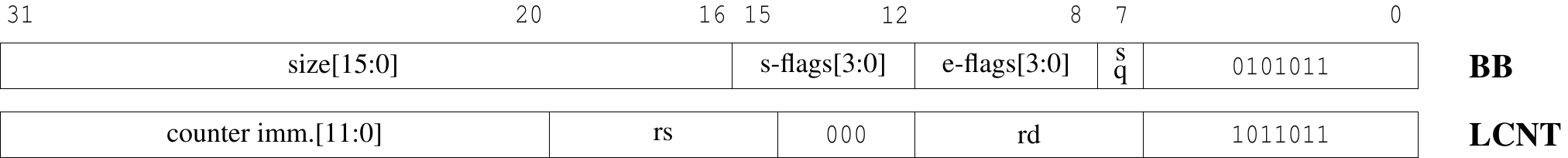}
\caption{\label{fig:riscv_instructions}Bitmap of the new RISC-V instructions.}
\end{figure*}

The
BasicBlocker
modification
requires
the
definition
of
the
\texttt{bb}
instruction
as
well
as
semantic
changes
to
all
control-flow
instructions.

The
\texttt{bb}
instruction
does
not
fit
into
any
of
the
existing
RISC-V
instruction
types
so
we
defined
a
new
instruction
type
to
achieve
an
optimal
utilization
of
the
instruction
bits
(Figure
\ref{fig:riscv_instructions}).
This
instruction
does
not
take
any
registers
as
input
but
rather
parses
the
information
directly
from
the
bitstring.
The
size
is
encoded
as
a
16-bit
immediate,
enabling
basic
blocks
with
up
to
65536
instructions.
One
can
split
a
larger
basic
block
into
multiple
sequential
blocks
if
necessary.
The
sequential
flag
is
a
one-bit
immediate
value.
The
behavior
of
all
RISC-V
control-flow
instructions
(\texttt{JAL},
\texttt{JALR},
\texttt{BEQ}, \texttt{BNE}, \texttt{BLT}, \texttt{BGE}, \texttt{BLTU}, \texttt{BGEU})
is
changed
so
that
they
alter
the
control
flow
at
the
end
of
the
current
basic
block.

We
also
include
hardware
loop
counters
in
the
BBRISC-V
ISA.
The
\texttt{lcnt}
instruction
sets
the
number
of
loop
iterations
(Figure
\ref{fig:riscv_instructions}). This I-Type instruction
requires
a
12
bit
immediate
value
as
well
as
a
source
and
a
target
register.
The
counter
value
is
then
computed
as
$cnt
=
imm+rs.value$
and
saved
to
the
loop
counter
set
defined
in
$rd$.
To
fully
support
loop
counters
we
also
add
four
start
and
end
flags
to
the
\texttt{bb}
instructions,
to
support
a
maximum
of
four
loop
counter
sets.

\subsection{CPU Implementation}
\subsubsection*{VexRiscv.}
For
the
soft
core
variant
of
an
in-order
CPU,
we
chose
the
32-bit
VexRiscv
core
\cite{vexriscv},
written
in
SpinalHDL.
This
soft
core
is
highly
configurable
by
the
use
of
plugins,
which
can
be
easily
extended
and
modified
to
include
new
functionalities.
We
use
a
configuration
with
five
stages
(IF,
ID,
EX,
MEM,
WB)
and
4096
byte,
one-way
instruction-
and
data
caches.
The
result
of
control-flow
instructions
is
available
after
the
memory
stage.
We
compare
the
modified
BasicBlocker
version
of
VexRiscv
against
the
original
core
with
the
best
available
branch
predictor
(dynamic
target).
To
enable
a
fair
comparison,
the
BasicBlocker
version
has
minimal
configuration
delta
to
the
original
core,
that
is
we
disabled
control-flow
speculation
and
added
the
logic
described
in
Section
\ref{sec:concept}.

Although
speculation
based
attacks
mostly
get
linked
to
out-of-order
CPUs
with
deep
pipelines,
they
are
also
feasible
on
smaller,
in-order
architectures~\cite{nemati2020speculative}
that
are
more
comparable
to
the
VexRiscv.

\subsubsection*{Gem5.}
To
simulate
the
performance
of
CPUs
with
superscalar
pipe\-lines
and
out-of-order
execution,
we
modified
the
64-bit
O3
CPU
model
of
the
Gem5
simulator~\cite{binkert2011gem5}.
The
Gem5
implementation
allows
high
configurability,
for
example
arbitrary
length
pipelines
can
easily
be
simulated
by
modifying
the
delays
between
two
stages.

In
the
default
configuration,
we
use
a
2x
superscalar
pipeline
configuration.
If
not
stated
otherwise,
we
use
the
default
configuration
supplied
in
the
\textit{se.py} configuration file. The simulated CPU is equipped with 64kB
L1
data
cache
and
32kB
instruction
cache.
Using
a
192
instruction
entry
sized
reorder
buffer,
the
CPU
can
execute
instructions
out-of-order.
As
for
the
VexRiscv
implementation,
the
BasicBlocker
version
makes
minimal
configuration
changes
to
enable
a
fair
comparison
of
performance
results.

\subsection{Compiler Modification}

To
be
able
to
evaluate
the
performance
of
our
concept
with
well
known
benchmark
programs
we
developed
a
compiler
supporting
and
optimizing
towards
our
instructions.
Our
compiler
is
based
on
the
LLVM
\cite{lattner2004llvm}
Compiler
Framework
version
10.0.0,
where
we
modified
the
RISC-V
backend
by
introducing
our
ISA
extension
and
inserting
new
compilation
passes
at
the
very
end
of
the
compilation
pipeline
to
not
interfere
with
other
passes
that
do
not
support
our
new
instructions.

First
of
all
we
split
basic
blocks
for
all
occurrences
of
call
instructions
since
they
break
the
consecutive
fetching
and
execution
of
instructions.
As
a
next
step
we
insert
the
\texttt{bb}
instructions
at
the
beginning
of
each
basic
block
that
include
the
number
of
instructions
in
the
block.
This
is
done
directly
before
code
emission
to
ensure
that
the
number
of
instructions
does
not
change
due
to
optimizations.
Linker
relaxation,
however,
is
one
optimization
that
could
reduce
the
number
of
instructions
by
substituting
calls
with
a
short
jumping
distance
by
a
single
jump
instruction
instead
of
two
instructions
(\texttt{aupic}
and
\texttt{jalr}).
Since
linker
relaxation
is
not
a
major
optimization,
we
simply
disabled
it,
but
it
would
also
be
possible
to
modify
the
linker
to
implement
BasicBlocker-aware
relaxation.

Our
modifications
to
the
semantics
of
terminating
instructions
(branches,
calls,
returns
and
jumps)
allow
them
to
be
scheduled
before
the
end
of
a
basic
block
and
rescheduling
them
earlier
is
also
crucial
to
the
performance
of
the
code.
This
is
done
in
a
top-down
list
scheduler
that
is
placed
after
register
allocation
and
prioritizes
terminating
instructions.
Additionally,
we
run
another
pass
afterwards
that
relocates
the
terminating
instructions
to
earlier
positions
in
the
basic
blocks
if
this
is
supported
by
register
dependencies.

\section{Evaluation}
\label{sec:eval}
In
the
following
we
provide
a
performance
evaluation
of
BasicBlocker
on
VexRiscv
and
Gem5
by
comparing
the
execution
time
of
different
variants
of
the
two
CPUs.
Thereby,
special
care
is
given
to
the
impact
of
CPU
features
and
code
characteristics.

\subsection{Selection of Benchmarks}
Both
implementations
of
BasicBlocker
presented
in
this
paper
enforce
the
presence
of
exactly
one
\texttt{bb}
instructions
in
every
basic
block
(i.e.
misplaced
or
missing
\texttt{bb}
instructions
cause
a
program
to
crash).
This
ensures
that
the
benchmarks
only
measure
the
performance
of
BasicBlocker
without
noise
from
legacy
code
snippets,
e.g.
library
functions,
but
also
requires
all
code
to
be
compiled
by
our
modified
compiler.
Since
this
forces
us
to
perform
the
benchmarks
bare-metal
(i.e.
without
OS
support),
it
is
quite
difficult
to
run
typical
user
level
benchmarks
such
as
SPEC.

We
chose
the
benchmarks
included
in
the
Embench
benchmark
suite
\cite{embenchIOT},
the
well-known
\textit{Coremark}
benchmark
\cite{gal2012exploring}
and
our
own
pointer-chasing
benchmark
for
our
evaluation.
The
selection
of
programs
within
the
Embench
suite
resemble
code
from
different
use
cases
such
as
cryptography
(\textit{nettle-sha},
\textit{nettle-aes}),
image
processing
(\textit{picojpeg})
and
matrix
multiplication
(\textit{matmult-int}).
For
three
of
the
programs
we
also
included
our
own
optimized
version
(\emph{-opt}),
targeted
at
general
architectures
and
discussed
in
more
detail
in
Appendix
\ref{algorithmic}.
All
those
programs
are
characterized
by
minimal
dependencies
and
are
thus
well
suited
for
bare-metal
benchmarking.

Since
all
of
the
benchmarks
require
the
\texttt{libc}
library
(and
some
also
\texttt{libm}), we compiled Newlib \cite{johnston2011newlib} using our modified LLVM compiler. However,
some
of
the
benchmark
programs
require
further
dependencies,
e.g.
\texttt{libgcc},
and
could
thus
not
be
compiled
for
our
target.
For
the
evaluation
we
included
all
available
benchmark
programs
that
compiled
with
the
modified
\texttt{libc}
and
\texttt{libm}
and
passed
the
test
for
functional
correctness.

We
compiled
three
versions
of
each
benchmark
program,
as
listed
in
Table~\ref{tab:versions}:
one
without
BasicBlocker,
one
with
a
new
compile
flag
enabling
the
insertion
of
\texttt{bb}
instructions,
and
one
with
\texttt{bb}
plus
rescheduling
of
terminator
instructions.
Except
for
these
differences,
the
compiler
and
compile
flags
are
identical.
The
compile
flags
are
listed
in
Appendix
\ref{appendix:cflags}.

\begin{table}[h]
\begin{center}
\begin{tabular}{ p{0.2\linewidth} || p{0.7\linewidth}  }
Name&Description\\
\hline\hline
\textit{Baseline} & Standard RISC-V version. \\
\textit{BB Info} & As in Baseline, but every basic block starts with a \texttt{bb} instruction.  \\
\textit{BB Resched} & As in BB Info, but with high-priority rescheduling of terminator instructions.
\end{tabular}
\end{center}
\caption{\label{tab:versions}Compiled versions used for benchmarking.}
\end{table}

We
ran
those
programs
on
several
variants
of
VexRiscv
and
Gem5,
as
listed
in
Table~\ref{tab:eval_versions}.
The
simplest
non-speculative
variant
(\textit{NoSpec})
disables
branch
prediction
and
speculative
fetching.
The
control-flow
speculation
configuration
(\textit{CFS})
implements
the
unmodified
version
of
the
CPU
with
the
default
branch
predictor.

\begin{table}[h]
\begin{center}
\begin{tabular}{p{0.5\linewidth} || p{0.08\linewidth} | p{0.08\linewidth} | p{0.1\linewidth}}
Name&BB&SF&BP
\\
\hline\hline
NoSpec
&
no
&
no
&
no\\
Control-Flow
Speculation
(CFS)
&
no
&
yes
&
yes\\
\textit{BasicBlocker} (this work) & yes & no & no
\end{tabular}
\end{center}
\caption{\label{tab:eval_versions}Processor instantiation options.
BB:
supports
{\tt
bb}
instruction.
SF:
speculative
fetching.
BP:
branch
predictor
}
\end{table}

As
we
execute
our
benchmarks
bare-metal,
we
observe
only
minimal
noise
through
the
microarchitectural
state
of
the
VexRiscv.
The
Gem5
platform
has
no
noise
at
all,
as
it
is
a
deterministic
simulation
with
a
reset
prior
to
each
run.
The
raw
benchmark
results
are
included
in
Appendix
\ref{appendix:rawBench}.

\subsection{VexRiscv Evaluation}

\begin{figure*}
\includestandalone[width=\textwidth]{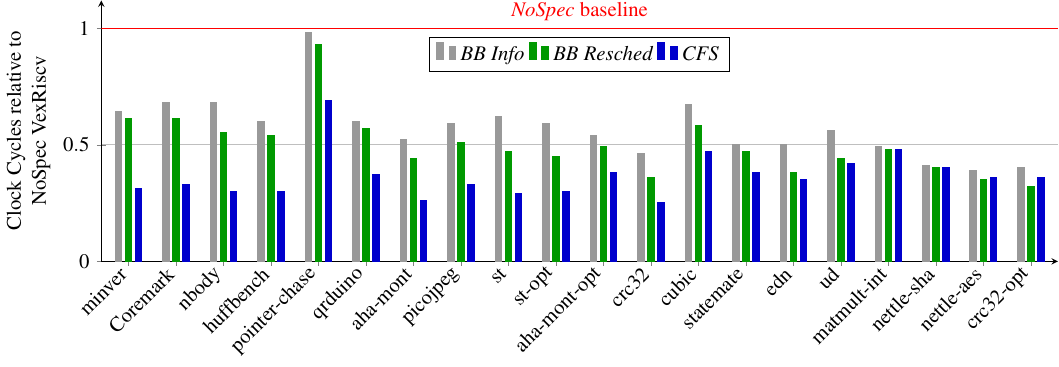}  	
\caption{\label{fig:benchmark_general_vex}Performance results for various benchmarks on VexRiscv measured in clock cycles.
The
results
are
relative
to
the
\textit{NoSpec}
configuration
of
VexRiscv
(red
line).
Sorted
descended
by
speedup
delta
in
\textit{BB
Resched}
vs
\textit{CFS}
case.
Lower
delta
is
better.
For
abbreviations
see
Tables~\ref{tab:versions}
and
\ref{tab:eval_versions}.
}
\end{figure*}

We
first
evaluate
the
performance
of
BasicBlocker
on
VexRiscv,
which
resembles
a
small-scale,
in-order,
embedded-like
processor,
by
comparing
the
execution
time
of
the
CPU
variants
in
Table
\ref{tab:eval_versions}
together
with
the
program
versions
of
Table
\ref{tab:versions}.
We
chose
the
strictly
non-control-flow-speculative
processor
as
a
naive
but
secure
baseline
and
report
the
relative
execution
time
of
the
other
variants
in
Figure
\ref{fig:benchmark_general_vex}.
The
average
speedup
over
all
benchmarks
is
$2.88\times$
and
$2.12\times$
for
the
version
using
control-flow
speculation
(\textit{CFS})
and
the
BasicBlocker
version
with
instruction
rescheduling
(\textit{BB
Resched}),
respectively.
The
maximal
and
minimal
speedups
are
$3.93\times$
(\emph{crc32})
and
$1.44\times$
(\emph{pointer-chase})
for
control
flow
speculation
and
$3.09\times$
(\emph{crc32-opt})
and
$1.07\times$
(\emph{pointer-chase})
for
BasicBlocker
with
rescheduling.

For
several
benchmarks
the
speedup
of
control-flow
speculation
is
comparable
to
BasicBlocker
with
instruction
rescheduling.
This
is
true
for
\emph{ud},
\emph{matmult-int},
\emph{nettle-sha},
\emph{nettle-aes},
and
\emph{crc32-opt}.
For
\emph{nettle-aes}
and
\emph{crc32-opt}
BasicBlocker
with
instruction
rescheduling
even
outperforms
control-flow
speculation
(speedup
of
$2.88\times$
vs.
$2.78\times$
and
$3.09\times$
vs.
$2.79\times$
respectively).
This
is
possible
as
with
enough
rescheduling
opportunities
no
pipeline
stalls
are
necessary
at
all.
For
other
benchmarks,
control-flow
speculation
outperforms
BasicBlocker
with
a
larger
margin
(e.g.
\emph{minver},
\emph{Coremark},
\emph{nbody},
and
\emph{huffbench}).

In
general,
BasicBlocker
performs
best
for
benchmarks
that
have
large
basic
blocks
and
less
branches
(e.g.
\emph{nettle-aes},
and
\emph{nettle-sha})
whereas
the
large
difference
of
speedup
between
control-flow
speculation
and
BasicBlocker
occurs
for
branch
heavy
code
with
small
basic
blocks
(e.g.
\emph{minver}).
A
more
thorough
analysis
of
code
characteristics
is
given
in
Section
\ref{sec:code_characteristics}.
We
emphasize
that
many
optimization
techniques
for
execution
time
tend
also
to
prefer
large
basic
blocks
with
less
branches
over
small
basic
blocks
with
a
lot
of
branches,
e.g.
loop
unrolling,
or
function
inlining.

\subsection{Gem5 Evaluation}

\begin{figure*}[h]
\includestandalone[width=\textwidth]{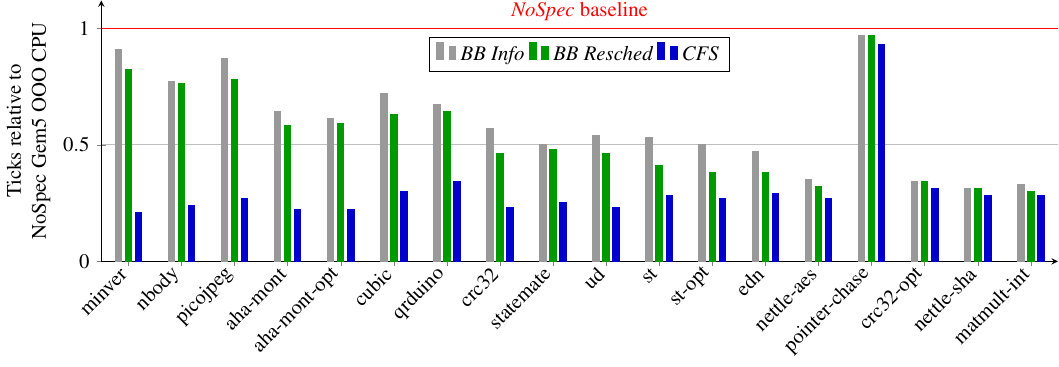}  	
\caption{\label{fig:benchmark_general_gem}Performance results for various benchmarks on Gem5 measured in simulation ticks.
The
results
are
relative
to
the
\textit{NoSpec}
configuration
of
Gem5
(red
line).
Sorted
descended
by
speedup
delta
in
\textit{BB
Resched}
vs
\textit{CFS}
case.
Lower
delta
is
better.
For
abbreviations
see
Tables~\ref{tab:versions}
and
\ref{tab:eval_versions}.
\textit{Huffbench} and \textit{Coremark} did not compile for the 64-bit target.}
\end{figure*}

We
conduct
the
same
performance
analysis
with
the
Gem5
simulator,
which
resembles
a
more
sophisticated,
out-of-order,
and
multi-scalar
processor.
Again,
the
strictly
non-control-flow-speculative
processor
variant
serves
as
a
naive
but
secure
baseline.
The
Gem5
CPU
model
processes
up
to
two
instructions
in
every
clock
cycle.
The
strictly
non-speculative
version
cannot
utilize
this
capacity
as
fetching
multiple
instructions
at
once
implies
speculative
fetching.
The
relative
execution
time
of
the
benchmarks
for
the
evaluated
processor
variants
are
reported
in
Figure
\ref{fig:benchmark_general_gem}.
The
average
speedup
over
all
running
benchmarks
is
$3.69\times$
and
$2.13\times$
for
the
version
using
control-flow
speculation
and
BasicBlocker
with
rescheduling
of
instruction
respectively.
The
maximum
and
minimum
speedups
are
$4.80\times$
(\emph{minver})
and
$1.07\times$
(\emph{pointer-chase})
for
control-flow
speculation
and
$3.09\times$
(\emph{crc32-opt})
and
$1.07\times$
(\emph{pointer-chase})
for
BasicBlocker
with
rescheduling.
Hence,
the
speedup
achieved
by
BasicBlocker
on
Gem5
is
overall
comparable
to
the
speedup
achieved
on
VexRiscv
and
for
well
performing
cases
slightly
higher.
However,
the
speedup
achieved
by
the
means
of
control-flow
speculation
is
higher
than
in
the
VexRiscv
example.

Taking
a
closer
look
at
specific
benchmarks
reveals
again
some
cases
where
BasicBlocker
matches
the
performance
of
control-flow
speculation,
e.g.
\emph{pointer-chase} \emph{crc32-opt}, \emph{nettle-sha}, or \emph{matmult-int} while for others control-flow speculation
is
considerably
faster,
e.g.
\emph{minver},
\emph{nbody},
or
\emph{picojpeg}.
As
analyzed
in
the
following,
the
code
characteristics
have
a
high
influence
on
the
performance.
The
low
speedup
for
\emph{pointer-chase}
at
all
Gem5
architectures
is
expected,
as
memory-access
time
clearly
dominates
any
pipeline
characteristic
for
this
benchmark.

The
results
show
the
applicability
of
BasicBlocker
on
superscalar,
out-of-order
processors.
We
further
analyze
the
influence
of
processor
characteristics
in
Section
\ref{sec:pipeline_characteristics}.

\subsection{Influence of Code Characteristics}
\label{sec:code_characteristics}
To
analyze
how
the
structure
of
the
code
influences
the
performance
of
BasicBlocker,
we
evaluate
the
code
characteristics
of
each
benchmark
regarding
the
average
size
of
basic
blocks
and
average
rescheduling
of
control-flow
instructions.
Since
the
impact
of
basic
blocks
that
are
executed
frequently
during
the
benchmarks
is
higher
than
those
that
are
executed
only
once,
we
perform
a
dynamic
hotspot
analysis
and
weight
the
results
based
on
the
frequency
of
invocation.
\begin{figure}[t]
\includestandalone[width=.95\linewidth]{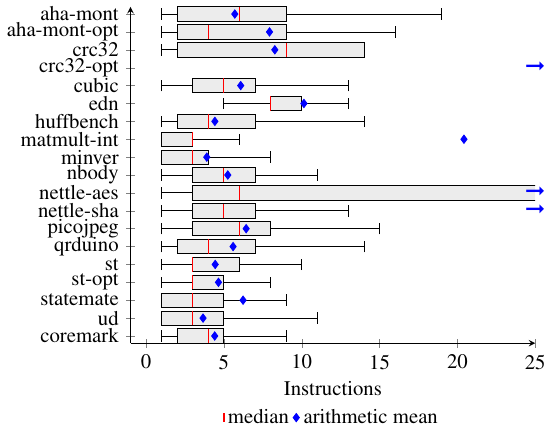}
\caption{\label{fig:basic_block_sizes}Distribution of basic block sizes (measured in instructions), weighted by the number of invocations, dynamically derived from the hotspot analysis.}
\end{figure}
In
Figure
\ref{fig:basic_block_sizes}
the
resulting
distribution
of
basic
block
sizes
is
pictured.
The
Figure
shows,
that
there
are
strong
differences
in
the
basic
block
sizes
for
the
benchmarks.
For
\textit{matmult-int},
\textit{nettle-aes}
and
\textit{nettle-sha}, the highest arithmetic average size of the basic blocks executed
during
the
benchmark
is
reached
with
more
than
25
instructions,
whereas
\textit{minver} and \textit{coremark} have a relatively small average basic
block
size,
below
five
instructions.
The
optimized
versions
of
\textit{aha-mont},
\textit{crc32}
and
\textit{st}
increase
the
mean
basic
block
size
by
enabling
more
inlining
and
thus
contribute
to
a
smaller
delta
in
the
benchmarks
between
the
BasicBlocker
and
speculative
version
of
the
cpu.
For
\textit{crc32-opt}
the
distribution
of
basic
block
sizes
changed
dramatically
and
lead
to
a
speedup
of
$2.13\times$
and
more
for
all
cpu
versions
compared
to
the
original
benchmark.

\begin{figure}[t]
\includestandalone[width=.95\linewidth]{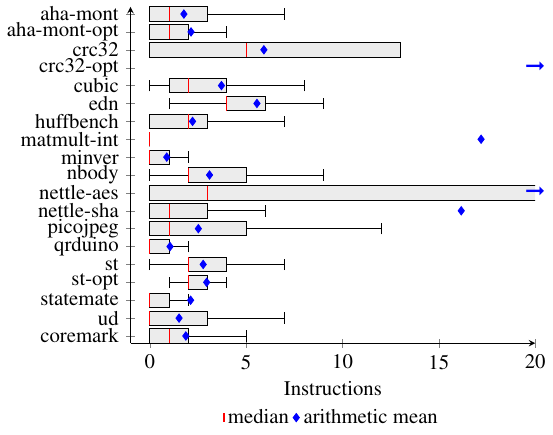}
\caption{\label{fig:basic_rescheduling}Distribution of instruction rescheduling per basic block, weighted by the number of invocations, dynamically derived from the hotspot analysis.}
\end{figure}

Figure
\ref{fig:basic_rescheduling}
shows
the
average
number
of
instructions
that
follow
the
control
flow
instruction
(this
is
only
relevant
for
the
\textit{BB Resched} case, not for the \textit{BB Info}). The intuitive assumption
is
that
large
basic
blocks
allow
for
higher
rescheduling
of
control
flow
instructions.
This
assumption
is
confirmed
by
the
results
shown
in
the
figure.
While
the
average
rescheduling
number
for
the
aforementioned
benchmarks
with
large
basic
blocks
is
high
(above
15
instructions
on
average),
benchmarks
with
smaller
basic
blocks
such
as
\textit{Coremark}
and
\textit{minver}
offer
less
average
rescheduling
opportunities.

The
performance
results
in
Figure
\ref{fig:benchmark_general_vex}
and
\ref{fig:benchmark_general_gem} show, that programs with large basic blocks
in
their
core
functions
(and
therefore
good
rescheduling
opportunities)
perform
better
with
BasicBlocker
than
those
benchmarks
with
small
basic
blocks.
For
real
world
workloads,
the
core
functions
that
are
regularly
executed
are
often
well
optimized
and
-
in
many
cases
-
try
to
avoid
branches
to
gain
improved
performance
\cite{elmasry2013branchless,elkhoulypattern,choi2001impact}.

\subsection{Influence of Pipeline Characteristics}
\label{sec:pipeline_characteristics}
\subsubsection*{Pipeline Length}
\begin{figure}[h]
\includestandalone[width=.8\linewidth]{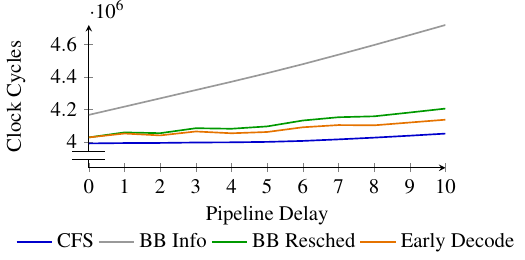}
\caption{\label{fig:vex_matmult_pipeline}Influence of additional pipeline stages on the execution time for the benchmark \emph{matmult-int} on VexRiscv.}
\end{figure}

\begin{figure}[h]
\includestandalone[width=.8\linewidth]{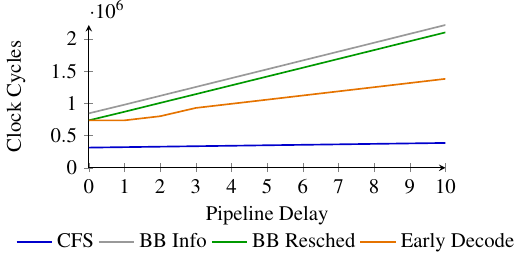}
\caption{\label{fig:vex_minver_pipeline}Influence of additional pipeline stages on the execution time for the benchmark \emph{minver} on VexRiscv.}
\end{figure}

We
analyze
the
influence
of
additional
pipeline
stages
on
the
execution
time
of
our
benchmarks
to
give
an
estimation
of
run
time
on
other
CPU
architectures.
As
for
space
restrictions
we
analyze
the
influence
of
the
pipeline
length
for
a
smaller
sample
of
the
above
shown
benchmarks.
With
\textit{matmult-int}
and
\textit{minver},
we
chose
one
well
performing
benchmark
and
one
with
higher
performance
penalty.
We
modified
the
VexRiscv
soft
core
and
placed
additional
dummy
pipeline
stages
between
fetch
and
decode
such
that
the
original
architecture
has
a
pipeline
delay
of
zero
and
each
additional
stage
increments
the
pipeline
delay
by
one.
The
results
are
shown
if
Figure
\ref{fig:vex_matmult_pipeline}
and
Figure
\ref{fig:vex_minver_pipeline}
for
\textit{matmult-int}
and
\textit{minver}
respectively.

The
data
clearly
show
that
additional
pipeline
stages
have
nearly
no
effect
when
control-flow
speculation
is
used
(CFS),
which
is
expected
as
the
longer
pipeline
only
introduces
a
penalty
if
a
missprediction
occurs.
Also
the
linearly
increasing
penalty
for
the
naive
BasicBlocker
implementation
is
to
be
expected,
since
a
constant
amount
of
additional
clock
cycles
is
added
to
all
transitions
between
basic
blocks.
More
interesting
is
the
case
where
the
compiler
is
allowed
to
reschedule
control-flow
instructions.
Here
we
can
see
clear
differences
between
the
benchmarks.
While
the
impact
of
additional
stages
is
only
small
and
non-linear
in
the
case
of
\emph{matmult-int}
running
on
VexRiscv,
we
can
observe
a
mirroring
of
the
naive
BasicBlocker
behavior
for
\emph{minver}
running
on
VexRiscv.
We
can
explain
this
as
an
artifact
of
the
code
structure,
as
discussed
earlier.
\emph{Minver}
is
composed
of
mostly
small
basic
blocks
resulting
in
only
a
few
rescheduling
options.
Hence,
the
impact
of
the
longer
pipeline
is
preserved
nearly
entirely.
In
contrast,
\emph{matmult-int}
has
better
options
for
rescheduling
and,
hence,
the
penalty
can
be
better
absorbed
through
the
early
determination
of
the
next
basic
block.

We
also
analyzed
one
additional
configuration,
where
we
implemented
a
decoding
of
the
\texttt{bb}
instruction
directly
after
the
instruction
cache
and,
hence,
before
the
pipeline
delay
is
introduced.
Figures
\ref{fig:vex_matmult_pipeline}
and
\ref{fig:vex_minver_pipeline}
show
that
this
can
reduce
the
performance
impact
of
longer
pipelines,
as
the
penalty
only
occurs
for
the
computation
of
the
next
basic
block
and
not
for
the
determination
of
the
basic
block
length
and
sequential
flag.

We
conducted
a
similar
analysis
for
the
Gem5
out-of-order
processor
and
the
results
show
the
same
behavior
as
the
discussed
examples,
as
can
be
seen
in
Appendix
\ref{appendix:gem5_pipeline}.

\subsubsection*{Superscalarity}
\begin{figure}[h]
\includestandalone[width=.8\linewidth]{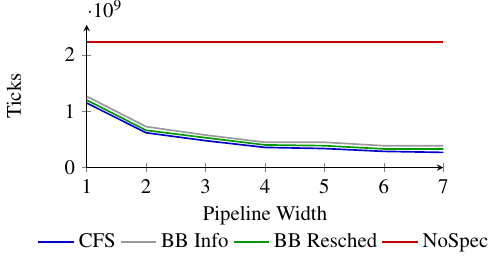}
\caption{\label{fig:gem_matmult_width}Influence of superscalarity on the performance of BasicBlocker using the \textit{matmult-int} benchmark on Gem5.}
\end{figure}

\begin{figure}[h]
\includestandalone[width=.8\linewidth]{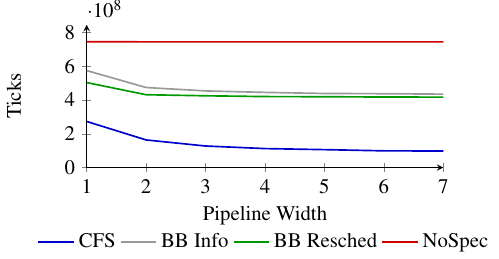}
\caption{\label{fig:gem_minver_width}Influence of superscalarity on the performance of BasicBlocker using the \textit{minver} benchmark on Gem5.}
\end{figure}

By
using
superscalarity
modern
processors
can
process
several
instructions
in
parallel
within
a
single
clock
cycle.
We,
therefore,
modify
our
Gem5
implementation
to
evaluate
the
performance
impact
of
superscalar
processors
using
BasicBlocker.
As
described
above,
our
default
configuration
for
the
Gem5
uses
a
$2\times$
superscalar
pipeline.
Figure
\ref{fig:gem_matmult_width}
and
\ref{fig:gem_minver_width}
show
the
performance
results
for
an
up
to
$7\times$
superscalar
pipeline
for
\textit{matmult-int}
and
\textit{minver}
respectively.
Graphs
for
other
benchmarks
can
be
found
in
Appendix
\ref{appendix:gem5_pipeline}.

The
red
line
in
Figures
\ref{fig:gem_matmult_width}
and
\ref{fig:gem_minver_width}
show
the
strictly
non-speculative
version
of
the
CPU.
Since
it
is
not
allowed
to
do
speculative
fetching,
only
one
instruction
can
be
fetched
at
a
time.
Thus,
the
superscalarity
has
no
effect
in
this
scenario.
The
results
for
the
well-performing
\textit{matmult-int}
benchmark
strikingly
demonstrate
the
potential
of
BasicBlocker
using
superscalar
pipelines.
The
\textit{bb
info}
version
as
well
as
the
\textit{rescheduled}
version
incur
minimal
performance
overhead
over
the
original
configuration
using
speculation.
That
is,
large
basic
blocks
allow
optimal
utilization
of
the
superscalar
pipeline.
For
the
\textit{minver} benchmark, which has much smaller basic blocks, it shows that the
additional
pipeline
slots
can
barely
be
filled
for
a
superscalarity
larger
than
two.
The
lines
for
\textit{bb
info}
and
\textit{rescheduled}
converge
for
a
large
pipeline
width.
That
is,
small
basic
blocks
will
eventually
be
fetched
within
a
single
clock
cycle,
making
any
rescheduling
irrelevant
to
the
performance.

\section{Conclusion}
\label{sec:conclusion}
In
this
work,
we
demonstrated
a
universal
countermeasure
against
control-flow
speculation
attacks
such
as
Spectre.
We
have
chosen
a
path
of
conservative
security
assumptions
that
completely
address
a
large
number
of
current
and
upcoming
attacks.
BasicBlocker
dispels
the
widely
accepted
assumption
that
control
flow
speculation
is
inevitable
for
performance.

We
propose
a
novel
concept
to
transport
control-flow
information
from
the
software
to
the
hardware,
enabling
practical
implementations
of
strictly
non-control-flow-speculative
processors.
The
performance
evaluation
clearly
shows
that
BasicBlocker
maintains
current
levels
of
performance
for
code
with
large
basic
blocks,
a
characteristic
that
is
common
in
highly
optimized
code
(i.e.
function
inlining,
loop
unrolling).
For
branch-heavy
code
control-flow
speculation
is
clearly
faster,
however,
this
is
at
the
cost
of
security.

In
contrast
to
other
work,
BasicBlocker
allows
to
remove
control-flow
speculation,
including
speculative-fetching,
entirely
and,
hence,
tackles
speculation-based
attacks
at
the
root
cause.
This
simplifies
the
security
analysis
drastically,
is
securely
backwards
compatible,
and
the
resulting
code
is
independent
of
the
underlying
microarchitecture.

We
showcase
our
concept
by
specifying
the
BBRISC-V
ISA,
including
a
concrete
implementation
of
that
ISA
based
on
VexRiscv
and
Gem5,
accompanied
by
an
optimizing
compiler
that
rests
on
the
LLVM
Compiler
Framework.
We
emphasize
that
BasicBlocker
is
a
generic
solution
that
can
be
applied
to
other
ISAs
as
well.
Our
prototype
implementations
show
that
BasicBlocker
is
applicable
for
a
variety
of
processor
types
and
we
point
to
code-optimization
strategies,
that
can
further
enhance
the
performance.

By
taking
the
algorithmic
level
into
consideration
further
optimizations
can
be
achieved;
see
Appendix~\ref{algorithmic}.
In
addition,
we
expect
extensions
and
future
work
to
improve
the
performance
and
security
of
BasicBlocker,
most
notably
hardware
loop
counters,
that
can
be
seamlessly
integrated
into
our
concept
(see
Appendix
\ref{appendix:loopCounter}),
or
extensions
dealing
with
fault-based
transient-execution
attacks.

\begin{acks}
The
authors
would
like
to
thank
Bastian
Kuttig
for
his
support
on
the
Gem5
evaluation.
Funded
by
the
Deutsche
Forschungsgemeinschaft
(DFG,
German
Research
Foundation)
under
Germany's
Excellence
Strategy
-
EXC
2092
CASA
-
390781972;
by
the
DFG
under
the
Priority
Program
SPP
2253
Nano
Security
(Project
RAINCOAT
-
Number:
440059533);
by
the
Cisco
University
Research
Program;
and
by
the
U.S.
National
Science
Foundation
under
grant
1913167.
"Any
opinions,
findings,
and
conclusions
or
recommendations
expressed
in
this
material
are
those
of
the
author(s)
and
do
not
necessarily
reflect
the
views
of
the
National
Science
Foundation"
(or
other
funding
agencies).
Date
of
this
document:
04
May
2021.
\end{acks}

\bibliographystyle{ACM-Reference-Format}

\begin{appendices}
\section{Loop Counter}
\label{appendix:loopCounter}

Loops
are
often
the
execution
hotspots
in
programs
and
contribute
considerably
to
diverging
control
flow.
Therefore
the
concept
of
hardware
supported
loops
can
be
profitable
as
already
discussed
in
the
literature
\cite{dipasquale2003hardware,
raghavan2008distributed}
and
implemented
in
various
architectures.

In
general,
hardware
loop
counters
are
realized
by
a
hardware
counter
which
is
set
by
a
dedicated
instruction
with
a
value
representing
the
maximum
trip
count
for
the
loop.
The
trip
count
must
be
computable
at
compile
time
to
be
inserted
by
an
immediate
value
or
available
in
a
register
at
run-time
before
entering
the
loop.
Information
about
which
instructions
are
included
in
the
loop
is
expressed
via
labels
or
additional
specific
instructions.
The
hardware
loop
counter
decrements
the
start
value
after
each
iteration
and
induces
a
branch
back
to
the
start
of
the
loop
as
long
as
the
counter
is
unequal
to
zero.
This
can
be
done
implicitly
at
the
end
of
the
loop
or
explicitly
with
an
instruction.

Performance
improvements
by
the
usage
of
hardware
loops
result
from
reduced
instruction
size
and
dedicated
loop
control
logic
that
does
not
have
to
be
calculated
by
the
ALU.
For
our
BasicBlocker
concept,
hardware
loops
are
actually
much
more
valuable
for
performance
when
only
applied
to
loops
that
will
not
terminate
early,
because
in
this
case
the
control
flow
for
all
loop
iterations
is
known
when
entering
the
loop.

We
can
seamlessly
support
hardware
loop
counters
in
our
design
concept,
by
introducing
a
new
instruction
and
adding
two
arguments
to
the
\texttt{bb}
instruction.
The
\texttt{lcnt} sets the number of loop iterations by storing a specified
value
into
a
dedicated
register.
The
start
and
end
address
of
the
loop
are
encoded
into
the
\texttt{bb}
instruction,
by
indicating
with
two
separate
flags
whether
the
corresponding
basic
block
is
the
start
or
end
block
of
the
loop.
These
two
flags
in
the
\texttt{bb}
instruction
are
necessary
for
each
loop
counter
set,
which
means
that
the
\texttt{bb}
instruction
needs
$2n$
bits
to
support
$n$
loop
counter
sets.
\definecolor{line1}{HTML}{AA4465}
\definecolor{line2}{HTML}{EDF0DA}
\definecolor{line3}{HTML}{A89B8C}
\definecolor{line4}{HTML}{F0DFAD}
\definecolor{line5}{HTML}{8F5C38}
\definecolor{line6}{HTML}{2E3532}
\definecolor{line7}{HTML}{2f6690}
\begin{lstlisting}[escapechar=|, language=riscv, label=lst:loopctr, caption=Single basic block loop with 3 iterations in counter set 1; Colors correspond to the execution trace in \ref{lst:execorder}, captionpos=b, numbers=right,]
|\colorbox{line1!50}{\mathstrut\hspace{.03\linewidth}}| bb 2, 1, 00, 00 ; len = 2, seq = 1
|\colorbox{line2!50}{\mathstrut\hspace{.03\linewidth}}| add a0, a0, a1
|\colorbox{line3!50}{\mathstrut\hspace{.03\linewidth}}| lcnt 3, lc1 ; 3 iterations, set 1
|\colorbox{line4!50}{\mathstrut\hspace{.03\linewidth}}| bb 2, 0, 01, 01 ; loop start/end
|\colorbox{line5!50}{\mathstrut\hspace{.03\linewidth}}| add a1, a2, a2
|\colorbox{line6!50}{\mathstrut\hspace{.03\linewidth}}| mul a2, a1, a2
|\colorbox{line7!50}{\mathstrut\hspace{.03\linewidth}}| bb 7, 0, 00, 00 ; after loop
\end{lstlisting}

Listing
\ref{lst:loopctr}
shows
the
exemplary
use
of
the
hardware
loop
counter.
In
line
3,
the
counter
in
loop
set
\texttt{ls1}
is
initialized
to
3.
The
following
\texttt{bb} instruction has the start- and end flag for loop set 1 enabled which indicates
a
loop
that
starts
at
the
beginning
of
this
basic
block
and
stretches
until
the
end
of
the
same
basic
block.
Each
bit
in
the
flags
represents
one
loop
counter
set,
allowing
nested
loops
with
the
same
start-
or
end
address
and
nested
loops
sharing
the
same
basic
block
as
start
or
end.
It
is
possible
to
model
loops
that
stretch
across
multiple
basic
blocks
by
setting
the
start
and
end
flags
in
the
respective
basic
blocks
accordingly.
When
the
\texttt{bb}
instruction
with
the
start
flag
is
executed,
the
current
$PC$
is
saved
as
start
address
in
the
corresponding
loop
counter
set.
Simultaneously,
the
counter
value
of
that
set
is
decremented
by
one.
When
the
execution
reaches
the
\texttt{bb} instruction with the corresponding end flag, the target address (which
determines
where
the
CPU
continues
execution)
is
set
to
the
corresponding
start
address
if
the
counter
is
not
zero.
Otherwise,
the
basic
block
is
handled
like
a
normal
sequential
basic
block
and
the
loop
will
exit.

\begin{lstlisting}[escapechar=|, language=riscv, label=lst:execorder, caption=Execution trace of CPU with color matched instructions to the code sequence in \ref{lst:loopctr}., captionpos=b]
|\colorbox{line1!50}{\rlap{bb 2, 1, 00, 00 ;  len = 2, seq. block}\hspace{\linewidth}\hspace{-2\fboxsep}}|
|\colorbox{line4!50}{\rlap{bb 2, 0, 01, 01 ; loop: start L1, end L1}\hspace{\linewidth}\hspace{-2\fboxsep}}|
|\colorbox{line2!50}{\rlap{add a0, a0, a1}\hspace{\linewidth}\hspace{-2\fboxsep}}|
|\colorbox{line3!50}{\rlap{lcnt 2, lc1 ; 2 iterations, set 1}\hspace{\linewidth}\hspace{-2\fboxsep}}|
|\colorbox{line4!50}{\rlap{bb 2, 0, 01, 01 ; loop: start L1, end L1}\hspace{\linewidth}\hspace{-2\fboxsep}}|
|\colorbox{line5!50}{\rlap{add a1, a2, a2}\hspace{\linewidth}\hspace{-2\fboxsep}}|
|\colorbox{line6!50}{\rlap{mul a2, a1, a2}\hspace{\linewidth}\hspace{-2\fboxsep}}|
|\colorbox{line4!50}{\rlap{bb 2, 0, 01, 01 ; loop: start L1, end L1}\hspace{\linewidth}\hspace{-2\fboxsep}}|
|\colorbox{line5!50}{\rlap{add a1, a2, a2}\hspace{\linewidth}\hspace{-2\fboxsep}}|
|\colorbox{line6!50}{\rlap{mul a2, a1, a2}\hspace{\linewidth}\hspace{-2\fboxsep}}|
|\colorbox{line7!50}{\rlap{bb 7, 0, 00, 00 ; after loop}\hspace{\linewidth}\hspace{-2\fboxsep}}|
|\colorbox{line5!50}{\rlap{add a1, a2, a2}\hspace{\linewidth}\hspace{-2\fboxsep}}|
|\colorbox{line6!50}{\rlap{mul a2, a1, a2}\hspace{\linewidth}\hspace{-2\fboxsep}}|
\end{lstlisting}

In
Listing
\ref{lst:execorder},
the
instruction
trace
of
the
program
snippet
from
Listing
\ref{lst:loopctr}
is
shown
as
it
is
executed
by
the
CPU.
Since
the
first
\texttt{bb} instruction indicates a sequential basic block, the CPU immediately
fetches
the
\texttt{bb}
instruction
of
the
next
basic
block
which
notifies
the
fetch
unit
that
the
second
basic
block
is
the
start
and
end
block
of
the
loop.
After
that,
the
remaining
\texttt{add}
and
\texttt{lcnt}
instructions
are
executed
to
finish
the
first
basic
block.
From
now
on
the
loop
counter
determines
the
execution
flow.
Since
the
second
basic
block
is
the
only
basic
block
of
the
loop,
the
\texttt{bb} instruction of this block is fetched again, to prepare the second loop round,
before
the
basic
block
is
executed
to
complete
the
first
round.
This
happens
again
until
the
loop
counter
is
zero,
resulting
in
fetching
the
last
\texttt{bb}
instruction,
to
exit
the
loop,
before
the
last
round
of
the
loop
is
executed.
Afterwards
the
execution
continues
outside
of
the
loop
with
the
normal
instruction
flow.

We
Implemented
our
proposed
hardware
loop
counter
concept
in
the
VexRiscv
core
and
added
elementary
compiler
support
for
one
loop
counter
set.
Because
the
loop
counter
can
only
be
used
for
loops
that
do
not
contain
calls
and
have
a
fixed
trip
count,
it
can
only
be
applied
by
the
compiler
to
a
small
subset
of
the
loops
in
the
benchmarks.
While
the
impact
of
the
hardware
loop
counter
is
neglegtable
for
most
benchmarks,
it
substantially
improves
the
speed
on
others.
The
speedup
for
\emph{edn}
improved
from
$2.63\times$
to
$2.70\times$,
getting
closer
to
the
$2.85\times$
speedup
of
the
speculative
version
compared
to
the
non-speculative
baseline.
For
\emph{ud}
the
hardware
loop
counter
enabled
the
BasicBlocker
variant
to
match
the
speed
of
the
speculative
version.
The
biggest
impact
can
be
observed
for
\emph{aha-mont}
where
the
speedup
increased
from
$2.27\times$
to
$3.13\times$.

\section{Synergies between BasicBlocker and algorithmic improvements}\label{algorithmic}

There
are
continual
announcements
of
performance
improvements
in
software
packages
to
handle
computational
``hot
spots'',
such
as
the
inner
loops
in
audio/video
processing.
The
main
point
of
this
appendix
is
that
the
natural
pursuit
of
higher-speed
software
favors
BasicBlocker:
software
changes
that
improve
performance
on
current
non-BasicBlocker
CPUs
tend
to
produce
even
larger
improvements
on
BasicBlocker
CPUs.

\subsection{Dimensions of performance analysis}
The
performance
evaluation
in
Section~\ref{sec:eval}
focuses
on
measuring
the
impact
of
changing
(1)
an
existing
CPU
with
an
existing
compiler
to
(2)
a
BasicBlocker
CPU
with
a
BasicBlocker-aware
compiler.
Each
of
the
benchmarks
being
compiled
and
run---for
example,
the
{\tt
st}
software
in
the
middle
of
the
graphs
in
that
section---is
treated
as
being
set
in
stone.
There
is
no
effort
in
Section~\ref{sec:eval}
to
modify
{\tt
st}
for
better
performance,
whether
by
explicit
changes
in
the
{\tt
st}
code
or
by
additions
to
the
compiler's
built-in
optimizations
beyond
the
BasicBlocker
support
described
earlier.

This
appendix
instead
treats
the
software
as
a
third
variable
beyond
the
compiler
and
the
CPU,
reflecting
the
reality
that
software
evolves
for
the
pursuit
of
performance.
For
example,
we
modified
the
{\tt
st}
software
to
obtain
the
{\tt
st-opt}
software
described
below,
computing
the
same
results
as
{\tt
st}
at
higher
speed.
Our
goal
in
changing
{\tt
st}
to
{\tt
st-opt}
was
to
match
what
typical
programmers
familiar
with
performance
would
naturally
do
if
{\tt
st}
turned
out
to
be
a
bottleneck.
We
used
a
profiler
(specifically
{\tt
gcc
-pg})
to
see
bottlenecks
on
an
existing
CPU
(specifically
the
ARM
Cortex-A7
CPU
in
a
Raspberry
Pi
2),
inspected
the
software
to
identify
underlying
inefficiencies,
and
removed
those
inefficiencies,
while
retaining
portability.

We
selected
three
case
studies
for
these
software
modifications:
{\tt
st},
{\tt
aha-mont},
and
{\tt
crc32}.
We
were
aiming
here
for
a
spread
of
different
types
of
code.
Within
our
benchmarks,
{\tt
aha-mont}
is
at
the
worst
quartile
for
BasicBlocker,
while
{\tt
st}
and
{\tt
crc32}
are
slightly
better
than
median;
{\tt
st}
uses
floating-point
arithmetic,
while
{\tt
aha-mont}
and
{\tt
crc32}
do
not.

It
is
important
to
observe
that
our
modifications
remove
{\it
cross-platform\/}
inefficiencies.
Switching
from
{\tt
st},
{\tt
aha-mont},
and
{\tt
crc32}
to
{\tt
st-opt},
{\tt
aha-mont-opt},
and
{\tt
crc32-opt}
saves
time
on
current
CPUs.
The
same
changes
save
time
on
BasicBlocker---and,
as
our
measurements
show,
reduces
the
cost
of
BasicBlocker
compared
to
current
CPUs.
We
summarize
the
inefficiencies
below
for
each
case
study,
and
explain
why
the
benefits
for
BasicBlocker
should
not
be
viewed
as
a
surprise.

\subsection{From {\tt st} to {\tt st-opt}}
Embench
describes
{\tt
st}
as
a
``statistics''
benchmark.
The
benchmark
computes
basic
statistics
regarding
two
length-100
arrays
of
double-precision
floating-point
numbers:
the
sum,
mean,
variance,
and
standard
deviation
of
each
array,
and
the
correlation
of
the
two
arrays.

However,
profiling
immediately
shows
that
most
of
the
time
in
{\tt
st}
is
spent
initializing
the
arrays.
In
general,
Embench
does
not
partition
the
function
being
benchmarked
from
the
preparation
of
input
to
the
function.
In
the
case
of
{\tt
st},
what
is
benchmarked
is
a
main
loop
that
calls
{\tt
Initialize}
for
one
array,
computes
the
sum
etc.~for
that
array,
calls
{\tt
Initialize}
for
the
other
array,
computes
the
sum
etc.~for
that
array,
computes
the
correlation,
and
repeats.

Embench
describes
itself
as
measuring
solely
``{\it
real\/}
programs'',
so
presumably
it
is
intentional
that
the
initialization
is
measured.
This
means
that
removing
the
initialization
from
the
benchmarks,
for
example
by
precomputing
the
{\tt
st}
arrays
at
compile
time,
would
not
be
a
valid
optimization.
The
operation
being
benchmarked
includes
computing
the
arrays
from
scratch
and
then
computing
statistics
given
the
arrays.

The
{\tt
st}
code
includes
a
function
computing
sum
and
mean,
a
function
computing
variance
and
standard
deviation,
and
a
function
computing
correlation.
The
sum
computation
is
almost
a
textbook
loop
through
the
input
array,
except
that
each
iteration
says
{\tt
*Sum
+=
Array[i]},
reading
and
updating
the
function
output
via
a
pointer;
{\tt
st-opt}
instead
does
the
textbook
{\tt
sum
+=
Array[i]},
using
a
local
{\tt
sum}
variable,
followed
by
{\tt
*Sum
=
sum}
after
the
loop.
The
second
and
third
functions
similarly
follow
textbook
formulas,
but
the
third
function
computes
the
variance
and
standard
deviation
again,
repeating
essentially
the
code
from
the
second
function;
{\tt
st-opt}
instead
saves
these
extra
results
from
the
third
function
and
eliminates
the
redundant
second
function.
If
one
array
were
involved
in
multiple
correlations
then
it
would
be
more
efficient
to
cache
the
standard
deviation.

The
initialization
in
{\tt
st},
before
the
statistics
are
computed,
sets
position
{\tt
i}
in
the
array
to
{\tt
i
+
RandomInteger()/8095.0}.
Here
{\tt
RandomInteger}
is
an
ad-hoc
linear-congruential
random-number
generator
where
each
output
is
the
previous
output
times
133
plus
81
modulo
8095.
On
typical
CPUs,
the
(integer
and
floating-point)
divisions
by
8095
are
expensive
operations,
more
expensive
than
a
series
of
several
loads,
additions,
and
multiplications
used
in
the
subsequent
statistical
computations.
Floating-point
operations
are
particularly
expensive
on
CPUs
without
floating-point
instructions,
such
as
VexRiscv,
since
each
floating-point
operation
is
then
implemented
by
a
``soft
float''
library,
although
it
is
not
clear
how
important
this
is
as
a
benchmarking
scenario.

There
are
faster
ways
to
produce
better-distributed
floating-point
numbers
between
0
and
1,
but
internally
{\tt
st}
checks
for
known
answers
for
these
particular
numbers,
so
let's
assume
that
computing
these
not-very-random
arrays
is
part
of
the
requirement.
It
is
well
known
how
to
convert
integer
division
into
a
short
sequence
of
multiplications,
shifts,
etc.;
{\tt
st-opt}
reduces
modulo
8095
in
this
way.
It
then
multiplies
by
{\tt
1/8095.0},
rather
than
dividing
by
{\tt
8095.0};
this
can
round
differently,
but
such
small
differences
are
conventionally
accepted
as
floating-point
optimizations
and,
more
to
the
point,
are
accepted
by
the
internal
{\tt
st}
tests.
The
{\tt
RandomInteger()}
function
is
inlined,
with
its
intermediate
outputs
being
kept
in
a
local
variable
and
saved
after
the
loop.

Finally,
each
of
these
loops
is
marked
in
{\tt
st-opt}
with
an
explicit
{\tt
UNROLL(4)}
or
{\tt
UNROLL(2)},
where
{\tt
UNROLL}
uses
existing
compiler
features
to
control
the
amount
of
unrolling.
The
overall
increase
from
{\tt
st}
compiled
code
size
to
{\tt
st-opt}
compiled
code
size
is
negligible:
around
100
bytes,
depending
on
the
instruction
set.

Except
for
the
possibility
of
branches
inside
a
``soft
float''
library,
there
is
nothing
inherently
unpredictable
in
the
{\tt
st}
control
flow:
the
program
sweeps
sequentially
through
length-100
arrays,
performing
the
same
sequence
of
operations
in
each
iteration.
The
short
basic
blocks
that
we
measured
in
{\tt
st},
averaging
under
5
instructions
with
median
just
3
instructions,
are
an
artifact
of
easily
removable
inefficiencies
described
above
in
{\tt
st},
such
as
the
redundant
loops
recomputing
variances,
the
loop
constantly
calling
a
separate
{\tt
RandomInteger}
function,
and
failures
of
unrolling.
Some
of
the
other
speedups
described
above,
such
as
eliminating
various
RAM
accesses,
do
not
increase
basic-block
sizes---on
the
contrary,
eliminating
these
instructions
makes
some
basic
blocks
shorter---but
this
leaves
room
for
further
unrolling,
again
improving
performance
across
platforms.

\subsection{From {\tt aha-mont} to {\tt aha-mont-opt}}
Embench
describes
{\tt
aha-mont}
as
a
``Montgomery
multiplication''
benchmark.
Montgomery
multiplication
is
a
well-known
method
to
carry
out
integer
operations
modulo
a
specified
odd
modulus
$m$
without
using
divisions
by
$m$.
The
{\tt
aha-mont}
code
is
a
slightly
modified
version
of
a
snippet
from
Warren's
``Hacker's
Delight''
code
corpus,
which
is
archived
at
\url{https://web.archive.org/web/20190715012506/http://hackersdelight.org/hdcode.htm}.

Profiling
again
shows
that
most
of
the
time
in
the
benchmark
is
actually
taken
by
something
else:
65\%
of
the
{\tt
aha-mont}
time
is
spent
in
divisions
by
$m$,
and
another
25\%
is
spent
in
an
{\tt
xbinGCD}
function,
while
Montgomery
multiplication
takes
under
10\%.
The
reason
that
there
are
divisions
by
$m$,
when
the
point
of
Montgomery
multiplication
is
to
avoid
divisions,
is
that
Warren's
snippet
includes
a
main
routine
with
tests,
and
the
tests
use
divisions.

The
modulus
$m$
is
a
{\tt
uint64},
possibly
as
large
as
$2^{64}-1$.
The
division-by-$m$
function
{\tt
modul64}
takes
two
{\tt
uint64}
inputs
$x$
and
$y$,
where
$x<m$,
and
returns
the
remainder
when
the
$128$-bit
integer
$2^{64}x+y$
is
divided
by
$m$.
The
code,
assuming
that
the
compiler
does
not
support
a
{\tt
uint128}
type,
uses
$64$
iterations
of
doubling
$2^{64}x+y$
and
subtracting
$m$
from
$x$
if
$x\ge
m$,
while
taking
care
to
check
for
the
possibility
that
the
doubling
overflows.
Overall
each
iteration
of
the
main
division
loop
uses
several
{\tt
uint64}
operations.

A
minor
inefficiency
here
is
as
follows.
The
code
was
originally
developed
to
compute
not
just
the
remainder
but
also
the
quotient.
The
obvious
way
to
do
this
is
to
add
$1$
to
a
new
variable
$q$
if
$x\ge
m$,
and
double
$q$
on
each
loop.
The
original
code
does
better
by
observing
that
the
space
needed
for
$q$
after
$i$
iterations,
namely
$i$
bits,
matches
the
space
cleared
at
the
bottom
of
$y$,
so
one
can
simply
add
$1$
to
$y$
if
$x\ge
m$,
which
eliminates
the
extra
doubling
of
$q$
since
$y$
is
being
doubled
anyway.
However,
in
the
context
of
{\tt
aha-mont},
the
quotient
is
thrown
away,
so
the
addition
is
a
waste
of
time.
The
merging
of
$q$
into
$y$
means
that
this
dead-code
elimination
is
beyond
what
the
compiler
figures
out
automatically.

There
is,
however,
a
much
larger
inefficiency
in
this
division
code,
namely
the
branches.
The
branches
involved
in
counting
$64$
iterations
are
predictable
and
can
be
straightforwardly
reduced
by
unrolling,
but
the
branches
involved
in
comparing
$x$
to
$m$
are
not.
One
expects
$0.5$
mispredictions
per
loop;
on
Intel
CPUs,
for
example,
this
would
cost
several
extra
cycles
per
loop.

Faster
division
algorithms---including
algorithms
that
handle
multiple
bits
at
a
time,
branchless
algorithms,
and
algorithms
that
precompute
a
reciprocal
of
$m$---are
not
a
new
topic,
and
in
fact
one
can
already
find
more
options
for
divisions
in
Warren's
code
corpus.
We
took
the
last
option
from
that
corpus---the
fastest,
according
to
the
documentation---and
incorporated
it
into
{\tt
aha-mont-opt}.
We
would
expect
anyone
who
cares
about
the
performance
of
this
code
to
benchmark
several
options
and
take
the
fastest
option
for
the
target
platform,
the
same
way
that
the
Linux
kernel
automatically
benchmarks
several
{\tt
raid6}
algorithms
and
selects
the
fastest.
Note
that
there
was
no
reason
for
Warren
to
bother
with
this
speedup
of
{\it
tests\/}
inside
his
Montgomery
snippet;
for
the
{\tt
aha-mont}
benchmark,
however,
these
tests
dominate
the
CPU
time.

The
{\tt
xbinGCD}
function
has
even
larger
cross-platform
branch-prediction
problems
than
{\tt
modul64}.
The
goal
here
is
to
compute
the
inverse
of
$m$
modulo
$2^{64}$;
this
is
a
precomputation
step
needed
for
Montgomery
multiplication.
The
{\tt
xbinGCD}
function
handles
this
with
a
general-purpose
binary-gcd
algorithm,
as
the
name
suggests.
Again
there
is
literature
on
more
efficient
algorithms---faster
ways
to
compute
binary
gcd,
and,
more
to
the
point,
faster
ways
to
compute
inverses
modulo
powers
of
$2$.
The
inversion
code
inside
{\tt
aha-mont-opt}
uses
just
$5$
iterations
(again
from
Warren's
code
corpus!),
where
each
iteration
uses
$2$
multiplications
and
$1$
subtraction;
this
is
an
order
of
magnitude
faster
(on
a
Raspberry
Pi
2)
than
the
inversion
code
inside
{\tt
aha-mont}.

It
is
clear
that
more
work
on
{\tt
aha-mont-opt}
would
produce
even
better
results,
especially
on
$32$-bit
platforms,
where
it
is
well
known
that
high-precision
computations
should
be
expressed
in
terms
of
$32$-bit
integers
rather
than
$64$-bit
integers.
For
RISC-V,
the
basic
instruction
set
is
unusual
in
that
it
does
not
include
carries,
and
it
also
does
not
include
conditional
arithmetic,
so
a
compiler
writer
implementing
{\tt
uint64}
in
terms
of
$32$-bit
operations
will
naturally
resort
to
branches.
Increased
attention
to
RISC-V
optimization
will,
presumably,
spur
development
of
branchless
carryless
algorithms
for
common
sequences
of
$64$-bit
operations---improving
performance
of
$64$-bit
code
on
existing
$32$-bit
RISC-V
CPUs,
and
improving
performance
even
more
with
BasicBlocker.

\subsection{From {\tt crc32} to {\tt crc32-opt}}
Embench
describes
{\tt
crc32}
as
a
``CRC
error
checking
32b''
benchmark.
The
main
{\tt
crc32pseudo}
function
computes
a
32-bit
cyclic
redundancy
check
of
8192
bits
of
data.
The
conventional
way
to
compute
a
CRC
is
to
update
the
CRC
for
$b$
bits
of
data
at
a
time,
using
a
few
$32$-bit
logic/shift
operations
and
a
lookup
of
$32$
bits
in
a
$2^b$-entry
table.
Both
{\tt
crc32}
and
{\tt
crc32-opt}
use
$b=8$,
so
there
are
$1024$
iterations
of
updates.

Profiling
once
again
shows
that
most
of
the
time
is
spent
on
something
else.
For
{\tt
crc32},
like
{\tt
st},
most
of
the
time
is
spent
setting
up
the
data.
Again
there
is
an
ad-hoc
linear-congruential
random-number
generator,
this
time
producing
each
$32$-bit
seed
as
the
previous
$32$-bit
seed
times
$1103515245$
plus
$12345$
modulo
$2^{31}$,
and
returning
the
top
$16$
bits
of
the
seed
(between
$0$
and
$2^{15}-1$)
as
output.
The
bottom
$8$
bits
of
the
output
are
then
used
as
the
next
$b=8$
bits
of
input
data
for
the
CRC.

For
{\tt
st}
the
obvious
costs
in
initialization
were
integer
and
floating-point
divisions
by
$8095$.
For
{\tt
crc32},
there
are
no
floating-point
operations,
and
the
reduction
modulo
$2^{31}$
is
already
written
as
a
logic
operation.
Furthermore,
the
initialization
loop
in
{\tt
crc32}
is
already
merged
into
the
CRC
computation
loop,
rather
than
having
one
pass
through
an
array
to
write
data
followed
by
a
separate
pass
through
the
array
to
process
data.

However,
each
iteration
of
the
{\tt
crc32}
loop
calls
a
function
{\it
in
a
separate
file}
to
generate
a
random
number.
The
compiler
does
not
inline
the
function.
The
only
changes
from
{\tt
crc32}
to
{\tt
crc32-opt}
are
(1)
putting
the
random-number-generation
function
into
the
same
file
for
inlining
and
(2)
marking
the
main
loop
with
{\tt
UNROLL(4)}.
The
unrolling
increases
code
size,
while
the
inlining
reduces
code
size
since
unnecessary
function
prologs
and
epilogs
disappear;
both
changes
in
code
size
are
negligible.

Note
that
it
is
already
common
practice
for
any
short
function
in
C
and
C++,
such
as
a
function
generating
a
random
number,
to
be
defined
in
a
{\tt
.h}
file,
so
that
the
compiler
can
easily
inline
the
function.
There
is
also
increasing
use
of
compiler
features
for
``link-time
optimization'',
which
has
the
same
basic
goal.

\subsection{Patterns observed, and consequences for BasicBlocker}
In
each
of
these
case
studies,
many
of
the
inefficiencies
in
the
original
code
arise
directly
from
loop
overhead
(and,
analogously,
function-call
overhead
in
the
{\tt
crc32}
case).
Branch
prediction
does
not
magically
make
loop
overhead
(and
function-call
overhead)
disappear;
it
can
reduce
the
overhead,
but
extremely
short
loops
(and
functions)
are
generally
performance
problems
if
they
are
in
hot
spots.
The
standard
response
is
unrolling
(plus
inlining)
for
hot
spots,
saving
time
on
current
CPUs---and
saving
even
more
time
for
BasicBlocker.

Further
inefficiencies
were
handled
by
copy
elimination
(e.g.,
removing
the
repeated
reads
and
writes
of
{\tt
*Sum}),
strength
reduction
(e.g.,
replacing
divisions
by
$8095$
with
multiplications),
and
common-subexpression
elimination
(e.g.,
eliminating
the
repeated
computation
of
variance)---which
can
indirectly
{\it
increase\/}
branch
frequency
by
reducing
the
time
spent
on
arithmetic
operations
between
branches.
However,
having
fewer
instructions
in
a
loop
usually
allows
more
unrolling
for
the
same
code
size,
and
then
branch
frequency
drops
again.

BasicBlocker
avoids
all
hot-spot
stalls
if
each
hot-spot
branch
condition
can
be
computed
enough
cycles
ahead
of
the
branch
to
cover
the
pipeline
length.
The
obvious
way
to
find
computations
that
are
{\it
intrinsically\/}
bad
for
BasicBlocker,
rather
than
being
bad
as
a
result
of
easily
fixable
failures
of
unrolling
and
inlining,
is
to
look
for
computations
bottlenecked
by
one
data-dependent
branch
feeding
into
another
data-dependent
branch,
such
as
the
bit-by-bit
data-dependent
branches
in
{\tt
modul64}
and
{\tt
xbinGCD}
inside
{\tt
aha-mont}.
We
emphasize
that
these
computations
also
perform
poorly
on
existing
CPUs;
we
saved
time
across
platforms
by
replacing
these
algorithms
with
faster
algorithms.

These
case
studies
are
not
necessarily
representative.
Are
there
important
computations
where
the
{\it
fastest\/}
algorithms
involve
one
data-dependent
branch
after
another?
There
is
a
textbook
example
at
this
point,
namely
sorting
integer
arrays.
Embench
includes
a
{\tt
wikisort}
benchmark
(which
did
not
compile
for
our
target),
stably
sorting
400
$64$-bit
records,
where
each
record
has
a
$32$-bit
integer
key
used
for
sorting
and
$32$
bits
of
further
data.
The
algorithm
used
inside
{\tt
wikisort}
is
a
complicated
merge-sort
variant;
overall
{\tt
wikisort}
has
1117
lines,
several
kilobytes
of
compiled
code.

However,
the
textbook
picture
of
the
fastest
sorting
algorithms
has
been
challenged
by
the
recent
speed
records
in~\cite{djbsort}
for
sorting
various
types
of
arrays
on
Intel
CPUs.
The
software
in~\cite{djbsort}
has
no
data-dependent
branches.
For
a
size-400
array,
this
software
uses
a
completely
predictable
pattern
of
7199
comparators
(size-2
sorting
operations,
i.e.,
min-max
operations);
merge
sort,
heap
sort,
etc.~use
half
as
many
comparisons
but
in
an
unpredictable
pattern,
incurring
so
much
overhead
as
to
be
non-competitive.

This
raises
a
research
question:
exactly
how
far
is
{\tt
wikisort}
from
optimal
on
smaller
CPUs?
An
application
where
sorting
is
critical
will
select
the
fastest
sorting
routine
from
among
many
options---not
just
comparison-based
sorts
such
as
merge
sort
but
also
radix
sort,
sorting
networks,
etc.
The
time
taken
by
{\tt
wikisort}
is,
presumably,
an
overestimate
of
the
time
needed
for
the
same
task
on
current
CPUs,
and
an
even
more
severe
overestimate
of
the
time
needed
for
the
same
task
on
BasicBlocker
CPUs.

More
broadly,
algorithms
without
data-dependent
branches
are
an
essential
part
of
the
modern
software-optimization
picture
for
large
CPUs,
especially
because
of
the
role
of
these
algorithms
inside
vectorized
code.
This
does
not
imply
that
these
algorithms
have
the
same
importance
on
today's
smaller
CPUs,
but
in
any
case
they
are
among
the
options
available
for
small
and
large
BasicBlocker
CPUs.
Taking
advantage
of
this
software
flexibility
brings
BasicBlocker
CPUs
even
closer
to
current
CPUs
in
overall
performance.
	\FloatBarrier
	
\section{Compile Flags}
\label{appendix:cflags}
The
compile
flags
for
the
Coremark
benchmark
are
listed
in
\cref{tab:cflags_coremark}
(omitting
includes,
debug,
macros,
toolchain
paths
and
flags
enabling
\texttt{bb}
instructions).
\begin{figure}[ht]
\centering
\begin{tabular}{|p{.4\linewidth} | p{.545\linewidth}|}
\hline
\textbf{Flag} & \textbf{Description}\\\hline\hline
O3
&
Optimization
Level
3\\\hline
march=rv32im
&
32-Bit
RISC-V
with
IM
extensions\\\hline
mabi=ilp32
&
Calling
convention
and
memory
layout
\\\hline
target=riscv32-unknown-elf
&
Select
target
architecture
\\\hline
mno-relax
&
No
linker
relexation\\\hline
lc
&
Link
C
library\\\hline
nostartfiles
&
Do
not
use
standard
system
startup
files
when
linking.\\\hline
ffreestanding
&
Only
use
features
available
in
freestanding
environment.
\\\hline
\end{tabular}
\caption{\label{tab:cflags_coremark} Coremark compile flags.}
\end{figure}

The
compile
flags
for
the
Embench
benchmarks
are
listed
in
\cref{tab:cflags_embench}
(omitting
includes,
debug,
macros,
toolchain
paths
and
flags
enabling
\texttt{bb}
instructions).
\begin{figure}[ht]
\centering
\begin{tabular}{|p{.4\linewidth} | p{.545\linewidth}|}
\hline
\textbf{Flag} & \textbf{Description}\\\hline\hline
O3
&
Optimization
Level
3\\\hline
march=[rv32im/rv64imfd]
&
32-Bit
RISC-V
for
VexRiscv
and
64-bit
for
Gem5\\\hline
mabi=[ilp32/lp64d]
&
Calling
convention
and
memory
layout
\\\hline
target=riscv[32/64]-unknown-elf
&
Select
target
architecture
\\\hline
mno-relax
&
No
linker
relexation\\\hline
fno-strict-aliasing
&
Disable
strict
aliasing.
\\\hline
\end{tabular}
\caption{\label{tab:cflags_embench} Embench compile flags.}
\end{figure}

	\FloatBarrier
\section{Pipeline Evaluation Graphs}
\label{appendix:gem5_pipeline}
This
Appendix
lists
the
graphs
for
pipeline
prolongation
for
other
benchmarks.

\subsection{VexRiscv Pipeline Length}
\begin{figure*}[ht]
\begin{minipage}[b]{0.31\linewidth}
\centering
\includestandalone[width=\linewidth]{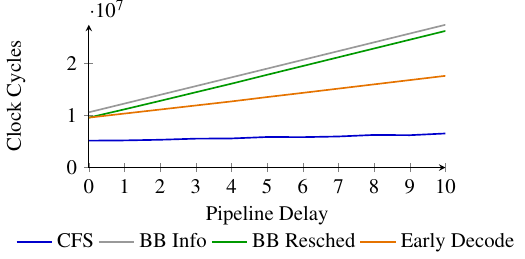}
\caption{\label{vex_start}VexRiscv - coremark}
\vspace{2ex}
\end{minipage}
\begin{minipage}[b]{0.31\linewidth}
\centering
\includestandalone[width=\linewidth]{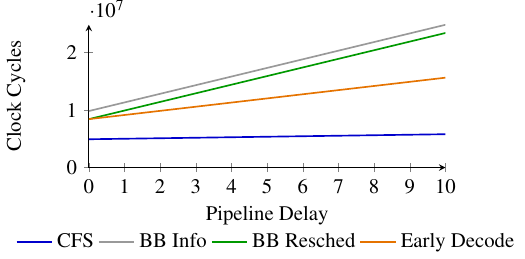}
\caption{VexRiscv - aha-mont}
\vspace{2ex}
\end{minipage}\begin{minipage}[b]{0.31\linewidth}
\centering
\includestandalone[width=\linewidth]{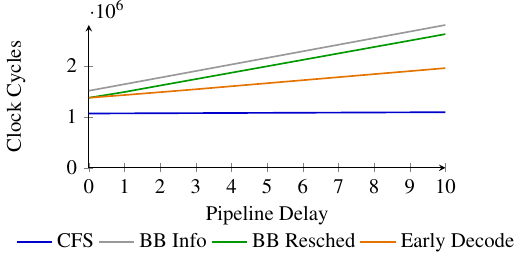}
\caption{VexRiscv - aha-mont-opt}
\vspace{2ex}
\end{minipage}
\begin{minipage}[b]{0.31\linewidth}
\centering
\includestandalone[width=\linewidth]{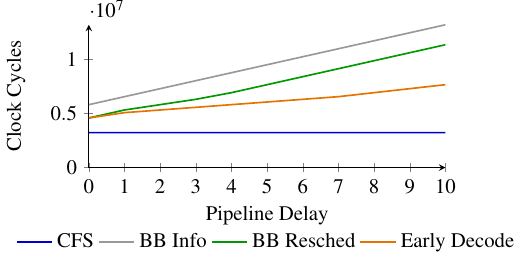}
\caption{VexRiscv - crc32}
\vspace{2ex}
\end{minipage}
\begin{minipage}[b]{0.31\linewidth}
\centering
\includestandalone[width=\linewidth]{tikzplots//benchmark_pipline_vex_crc32-opt}
\caption{VexRiscv - crc32-opt}
\vspace{2ex}
\end{minipage}
\begin{minipage}[b]{0.31\linewidth}
\centering
\includestandalone[width=\linewidth]{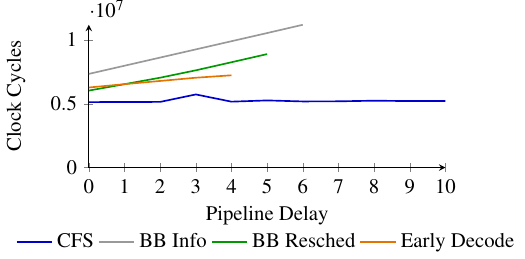}
\caption{VexRiscv - edn}
\vspace{2ex}
\end{minipage}
\begin{minipage}[b]{0.31\linewidth}
\centering
\includestandalone[width=\linewidth]{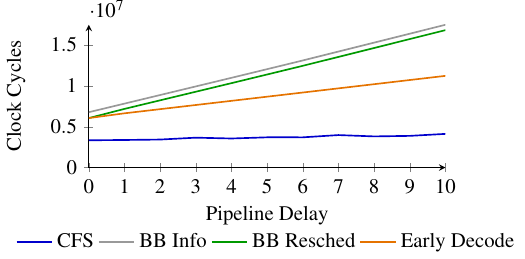}
\caption{VexRiscv - huffbench}
\vspace{2ex}
\end{minipage}
\begin{minipage}[b]{0.31\linewidth}
\centering
\includestandalone[width=\linewidth]{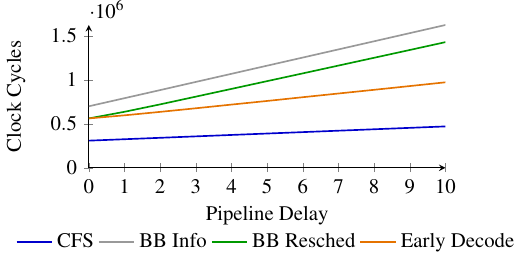}
\caption{VexRiscv - nbody}
\vspace{2ex}
\end{minipage}
\begin{minipage}[b]{0.31\linewidth}
\centering
\includestandalone[width=\linewidth]{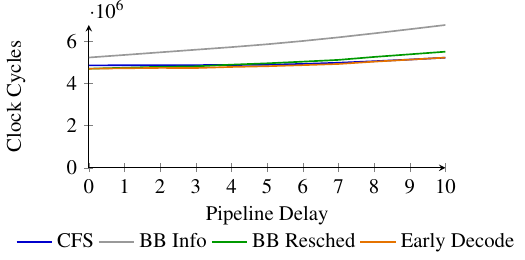}
\caption{VexRiscv - nettle-aes}
\vspace{2ex}
\end{minipage}
\begin{minipage}[b]{0.31\linewidth}
\centering
\includestandalone[width=\linewidth]{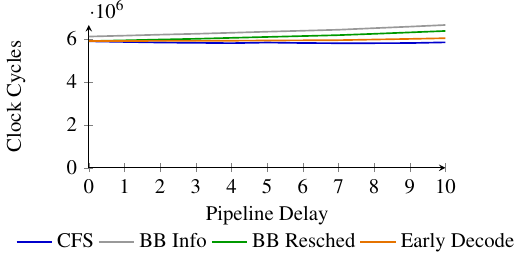}
\caption{VexRiscv - nettle-sha}
\vspace{2ex}
\end{minipage}
\begin{minipage}[b]{0.31\linewidth}
\centering
\includestandalone[width=\linewidth]{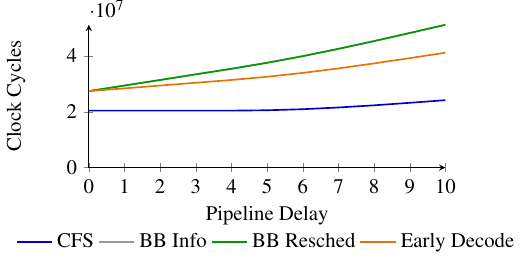}
\caption{VexRiscv - pointer-chase}
\vspace{2ex}
\end{minipage}
\begin{minipage}[b]{0.31\linewidth}
\centering
\includestandalone[width=\linewidth]{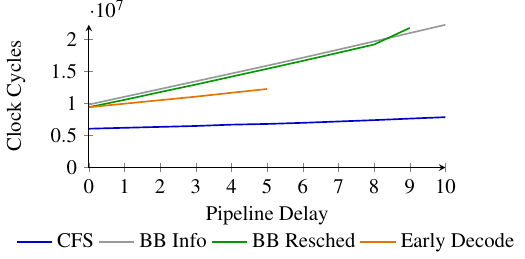}
\caption{VexRiscv - qrduino}
\vspace{2ex}
\end{minipage}
\begin{minipage}[b]{0.31\linewidth}
\centering
\includestandalone[width=\linewidth]{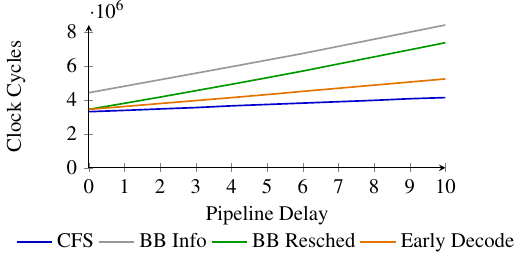}
\caption{VexRiscv - st}
\vspace{2ex}
\end{minipage}
\begin{minipage}[b]{0.31\linewidth}
\centering
\includestandalone[width=\linewidth]{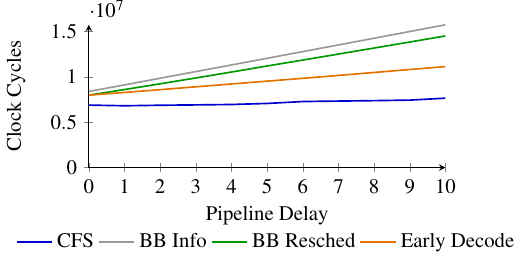}
\caption{VexRiscv - statemate}
\vspace{2ex}
\end{minipage}
\begin{minipage}[b]{0.31\linewidth}
\centering
\includestandalone[width=\linewidth]{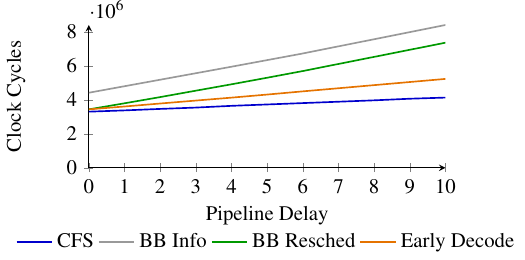}
\caption{\label{vex_end}VexRiscv - ud}
\vspace{2ex}
\end{minipage}

\end{figure*}

\Cref{vex_start} to \ref{vex_end} show the graphs for VexRiscv.
\renewcommand\figurename{Fig.}

\bigskip
\FloatBarrier

\subsection{Gem5 Pipeline Length}
\begin{figure*}[ht]
\begin{minipage}[b]{0.31\linewidth}
\centering
\includestandalone[width=\linewidth]{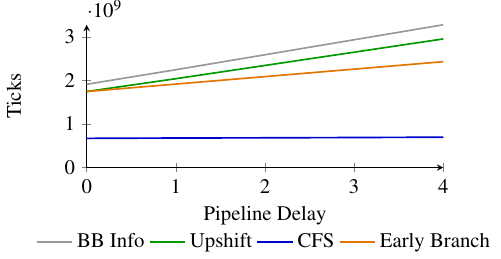}
\caption{\label{gem5_start}Gem5 - aha-mont}
\vspace{2ex}
\end{minipage}\begin{minipage}[b]{0.31\linewidth}
\centering
\includestandalone[width=\linewidth]{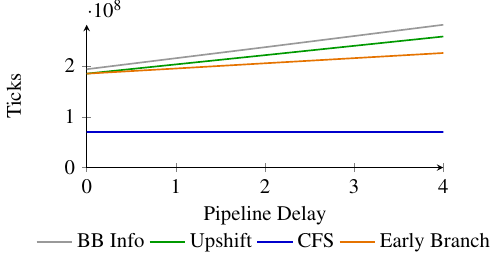}
\caption{Gem5 - aha-mont-opt}
\vspace{2ex}
\end{minipage}\begin{minipage}[b]{0.31\linewidth}
\centering
\includestandalone[width=\linewidth]{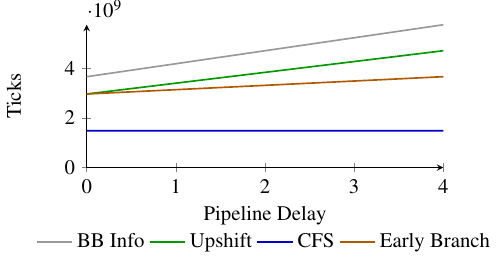}
\caption{Gem5 - crc32}
\vspace{2ex}
\end{minipage}
\begin{minipage}[b]{0.31\linewidth}
\centering
\includestandalone[width=\linewidth]{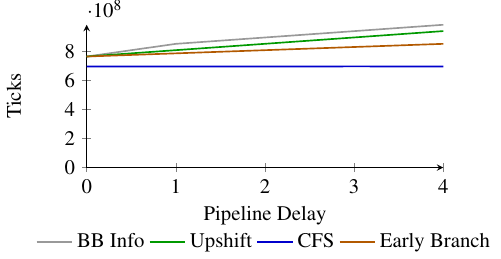}
\caption{Gem5 - crc32-opt}
\vspace{2ex}
\end{minipage}
\begin{minipage}[b]{0.31\linewidth}
\centering
\includestandalone[width=\linewidth]{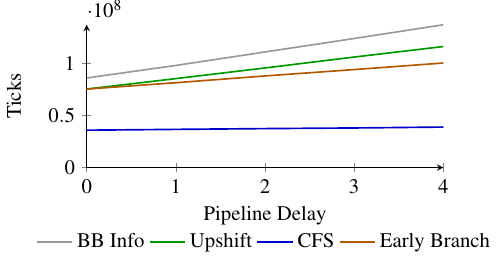}
\caption{Gem5 - cubic}
\vspace{2ex}
\end{minipage}
\begin{minipage}[b]{0.31\linewidth}
\centering
\includestandalone[width=\linewidth]{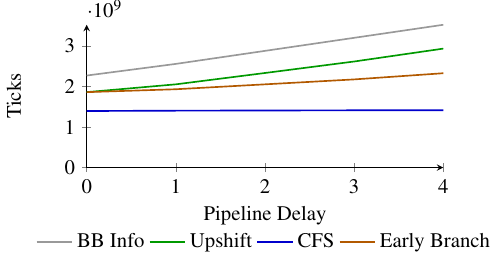}
\caption{Gem5 - edn}
\vspace{2ex}
\end{minipage}
\begin{minipage}[b]{0.31\linewidth}
\centering
\includestandalone[width=\linewidth]{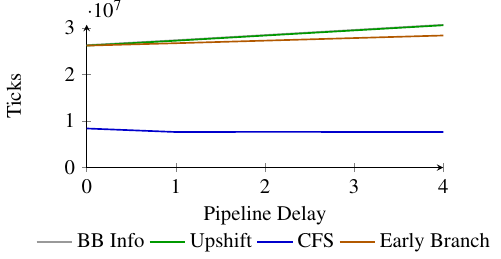}
\caption{Gem5 - nbody}
\vspace{2ex}
\end{minipage}
\begin{minipage}[b]{0.31\linewidth}
\centering
\includestandalone[width=\linewidth]{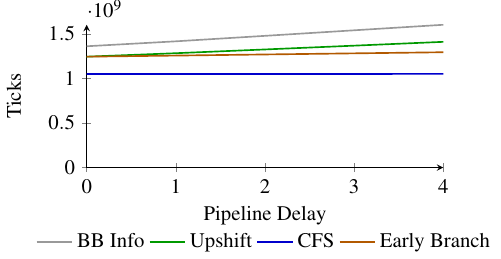}
\caption{Gem5 - nettle-aes}
\vspace{2ex}
\end{minipage}
\begin{minipage}[b]{0.31\linewidth}
\centering
\includestandalone[width=\linewidth]{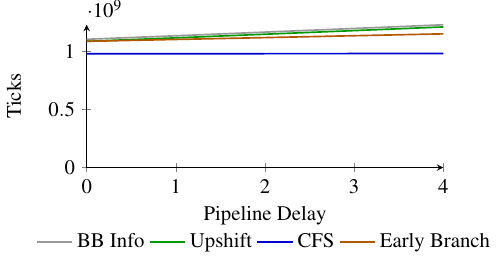}
\caption{Gem5 - nettle-sha}
\vspace{2ex}
\end{minipage}
\begin{minipage}[b]{0.31\linewidth}
\centering
\includestandalone[width=\linewidth]{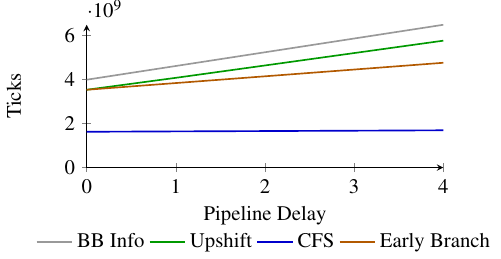}
\caption{Gem5 - picojpeg}
\vspace{2ex}
\end{minipage}
\begin{minipage}[b]{0.31\linewidth}
\centering
\includestandalone[width=\linewidth]{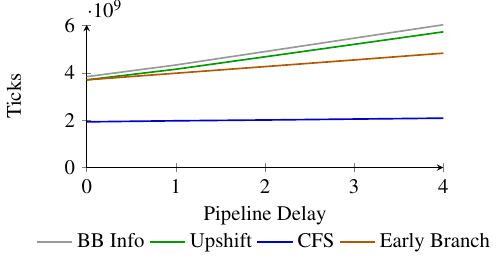}
\caption{Gem5 - qrduino}
\vspace{2ex}
\end{minipage}
\begin{minipage}[b]{0.31\linewidth}
\centering
\includestandalone[width=\linewidth]{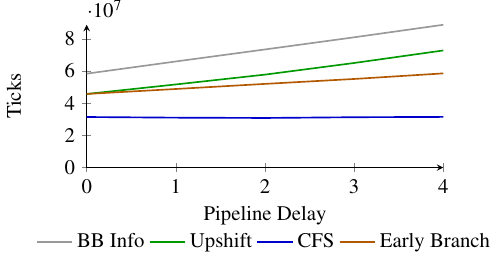}
\caption{Gem5 - st}
\vspace{2ex}
\end{minipage}
\begin{minipage}[b]{0.31\linewidth}
\centering
\includestandalone[width=\linewidth]{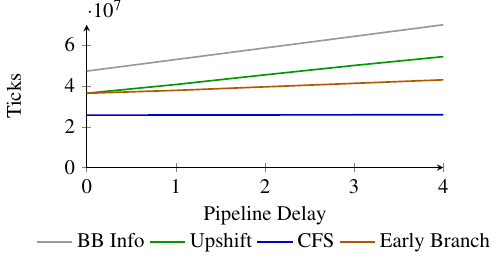}
\caption{Gem5 - st-opt}
\vspace{2ex}
\end{minipage}
\begin{minipage}[b]{0.31\linewidth}
\centering
\includestandalone[width=\linewidth]{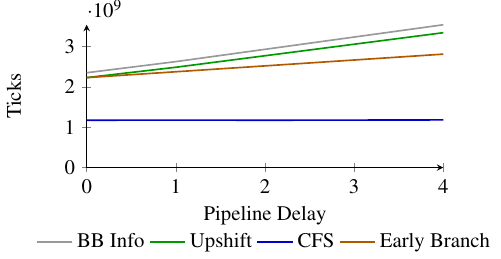}
\caption{Gem5 - statemate}
\vspace{2ex}
\end{minipage}
\begin{minipage}[b]{0.31\linewidth}
\centering
\includestandalone[width=\linewidth]{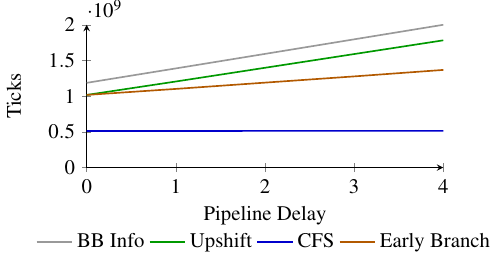}
\caption{\label{gem5_end}Gem5 - ud}
\vspace{2ex}
\end{minipage}
\end{figure*}

\Cref{gem5_start} to \ref{gem5_end} show the graphs for Gem5.

\bigskip
\FloatBarrier

\subsection{Gem5 Pipeline Width}
\begin{figure*}[ht]
\label{ fig7}
\begin{minipage}[b]{0.31\linewidth}
\centering
\includestandalone[width=\linewidth]{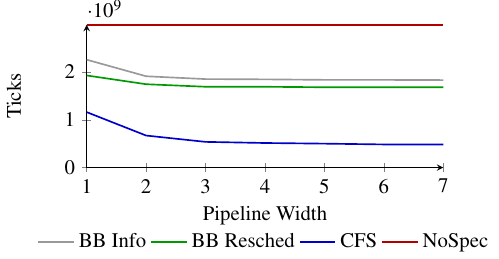}
\caption{\label{gem5_start2}Gem5 - aha-mont}
\vspace{2ex}
\end{minipage}\begin{minipage}[b]{0.31\linewidth}
\centering
\includestandalone[width=\linewidth]{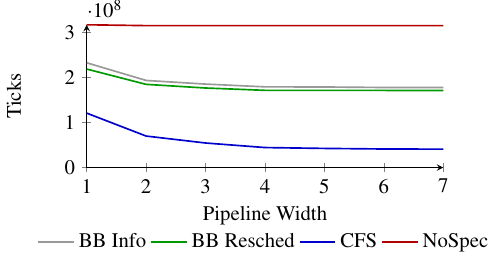}
\caption{Gem5 - aha-mont-opt}
\vspace{2ex}
\end{minipage}
\begin{minipage}[b]{0.31\linewidth}
\centering
\includestandalone[width=\linewidth]{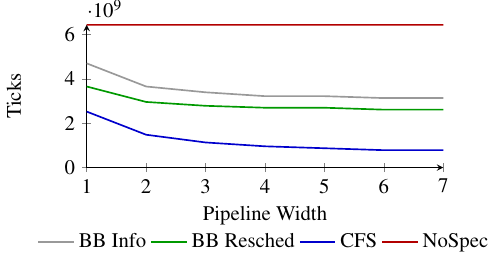}
\caption{Gem5 - crc32}
\vspace{2ex}
\end{minipage}
\begin{minipage}[b]{0.31\linewidth}
\centering
\includestandalone[width=\linewidth]{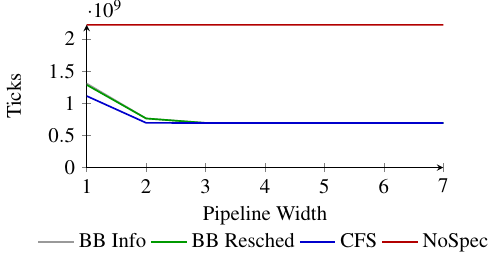}
\caption{Gem5 - crc32-opt}
\vspace{2ex}
\end{minipage}
\begin{minipage}[b]{0.31\linewidth}
\centering
\includestandalone[width=\linewidth]{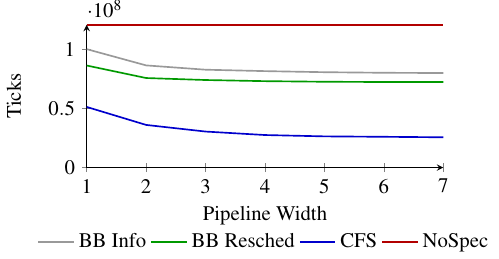}
\caption{Gem5 - cubic}
\vspace{2ex}
\end{minipage}
\begin{minipage}[b]{0.31\linewidth}
\centering
\includestandalone[width=\linewidth]{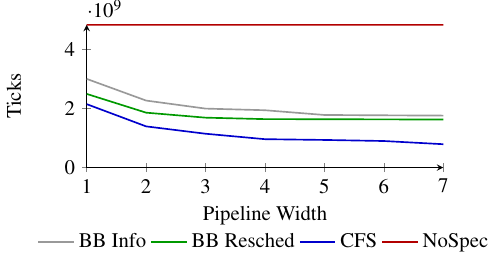}
\caption{Gem5 - edn}
\vspace{2ex}
\end{minipage}
\begin{minipage}[b]{0.31\linewidth}
\centering
\includestandalone[width=\linewidth]{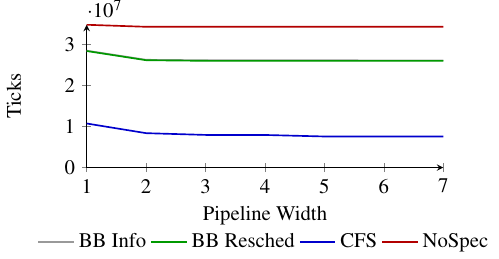}
\caption{Gem5 - nbody}
\vspace{2ex}
\end{minipage}
\begin{minipage}[b]{0.31\linewidth}
\centering
\includestandalone[width=\linewidth]{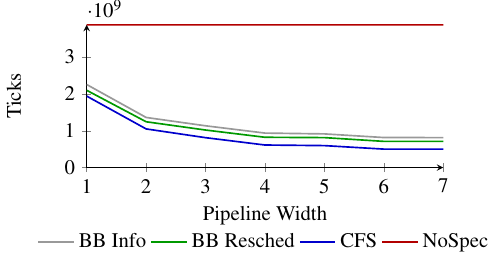}
\caption{Gem5 - nettle-aes}
\vspace{2ex}
\end{minipage}
\begin{minipage}[b]{0.31\linewidth}
\centering
\includestandalone[width=\linewidth]{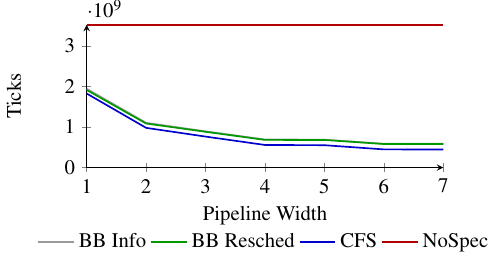}
\caption{Gem5 - nettle-sha}
\vspace{2ex}
\end{minipage}
\begin{minipage}[b]{0.31\linewidth}
\centering
\includestandalone[width=\linewidth]{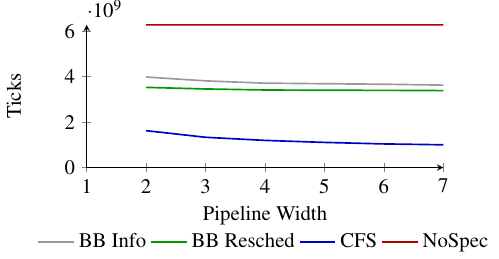}
\caption{Gem5 - picojpeg}
\vspace{2ex}
\end{minipage}
\begin{minipage}[b]{0.31\linewidth}
\centering
\includestandalone[width=\linewidth]{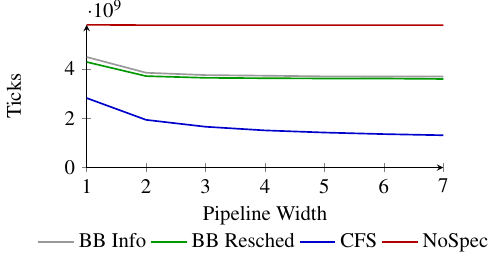}
\caption{Gem5 - qrduino}
\vspace{2ex}
\end{minipage}
\begin{minipage}[b]{0.31\linewidth}
\centering
\includestandalone[width=\linewidth]{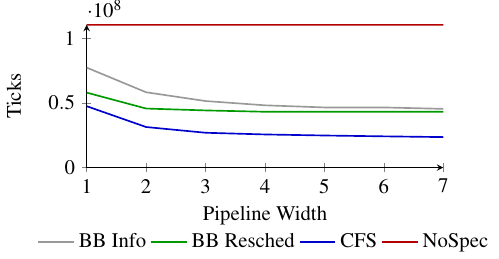}
\caption{Gem5 - st}
\vspace{2ex}
\end{minipage}
\begin{minipage}[b]{0.31\linewidth}
\centering
\includestandalone[width=\linewidth]{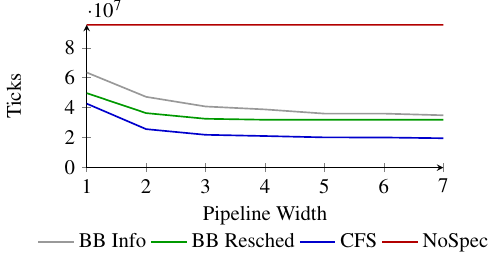}
\caption{Gem5 - st-opt}
\vspace{2ex}
\end{minipage}
\begin{minipage}[b]{0.31\linewidth}
\centering
\includestandalone[width=\linewidth]{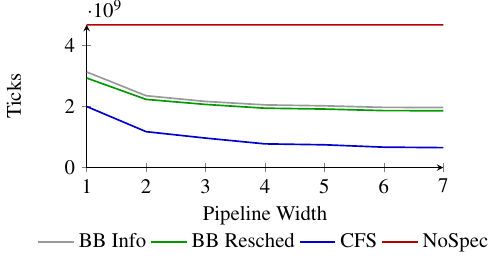}
\caption{Gem5 - statemate}
\vspace{2ex}
\end{minipage}
\begin{minipage}[b]{0.31\linewidth}
\centering
\includestandalone[width=\linewidth]{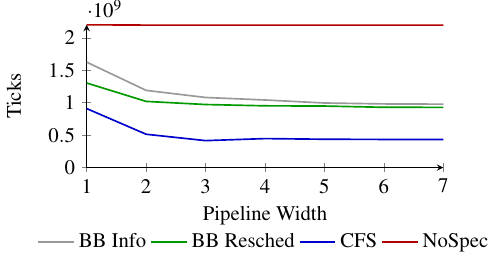}
\caption{\label{gem5_end2}Gem5 - ud}
\vspace{2ex}
\end{minipage}

\end{figure*}

\Cref{gem5_start2} to \ref{gem5_end2} show the graphs for Gem5.

	\FloatBarrier
\newcolumntype{P}[1]{>{\raggedleft\arraybackslash}p{#1}}
\begin{figure*}[htb!]
\centering
\resizebox{\textwidth}{!}{\begin{tabular}{ |p{.1\textwidth}||
P{.1\textwidth}|P{.1\textwidth}|P{.1\textwidth}||P{.1\textwidth}|P{.1\textwidth}|P{.1\textwidth}||P{.1\textwidth}|P{.1\textwidth}|P{.1\textwidth}||P{.1\textwidth}|P{.1\textwidth}|P{.1\textwidth}|
}
\hline
&\multicolumn{3}{|c||}{\textbf{Speculation}}&\multicolumn{3}{c|}{\textbf{No
Speculation}}&\multicolumn{3}{c|}{\textbf{BB
Info}}&\multicolumn{3}{c|}{\textbf{Rescheduling}}\\
\hline
\textbf{Benchmark} & \multicolumn{1}{|c|}{\textbf{Quartile}} & \multicolumn{1}{|c|}{\textbf{Median}} & \multicolumn{1}{|c||}{\textbf{Quartile}} & \multicolumn{1}{c|}{\textbf{Quartile}} & \multicolumn{1}{|c|}{\textbf{Median}} & \multicolumn{1}{|c|}{\textbf{Quartile}} & \multicolumn{1}{c|}{\textbf{Quartile}} & \multicolumn{1}{|c|}{\textbf{Median}} & \multicolumn{1}{|c|}{\textbf{Quartile}}& \multicolumn{1}{c|}{\textbf{Quartile}} & \multicolumn{1}{|c|}{\textbf{Median}} & \multicolumn{1}{|c|}{\textbf{Quartile}}\\
\hline\hline
Coremark&
5175116
&
5175286
&
5175482
&
15707821
&
15708006
&
15708178
&
10622955
&
10623282
&
10623472
&
9595350
&
9595555
&
9595848
\\\hline
aha-mont&
4932786
&
4932801
&
4932815
&
19002352
&
19002370
&
19002382
&
9839038
&
9839051
&
9839068
&
8430604
&
8430616
&
8430627
\\\hline
aha-mont-opt&
1066889
&
1066913
&
1066938
&
2808234
&
2808263
&
2808285
&
1513601
&
1513649
&
1513694
&
1375966
&
1376001
&
1376045
\\\hline
crc32&
3230872
&
3230948
&
3231032
&
12696521
&
12696615
&
12696721
&
5813490
&
5813632
&
5813764
&
4584968
&
4585061
&
4585196
\\\hline
crc32-opt&
2699898
&
2699900
&
2699902
&
7532941
&
7532943
&
7532947
&
3005250
&
3005253
&
3005256
&
2439483
&
2439486
&
2439489
\\\hline
cubic&
537960
&
538204
&
538317
&
1149501
&
1149668
&
1149844
&
773078
&
773310
&
773452
&
667308
&
667458
&
667704
\\\hline
edn&
5133587
&
5133694
&
5133840
&
14697842
&
14697951
&
14698042
&
7345958
&
7346067
&
7346179
&
5514993
&
5515155
&
5515294
\\\hline
huffbench&
3342980
&
3343182
&
3343359
&
11327279
&
11327477
&
11327654
&
6780333
&
6780629
&
6780821
&
6062285
&
6062506
&
6062668
\\\hline
matmult-int&
4478246
&
4478557
&
4478918
&
9328936
&
9329132
&
9329354
&
4611543
&
4611884
&
4612178
&
4470991
&
4471335
&
4471712
\\\hline
minver&
98144
&
98220
&
98281
&
316393
&
316494
&
316578
&
203192
&
203280
&
203358
&
194293
&
194372
&
194467
\\\hline
nbody&
308486
&
308528
&
308562
&
1025912
&
1025956
&
1026018
&
698678
&
698770
&
698878
&
560584
&
560717
&
560830
\\\hline
nettle-aes&
4862599
&
4863024
&
4863262
&
13528174
&
13528508
&
13528876
&
5236309
&
5236621
&
5236992
&
4703939
&
4704062
&
4704256
\\\hline
nettle-sha&
5879961
&
5881300
&
5881465
&
14817231
&
14817347
&
14817462
&
6099560
&
6100163
&
6100944
&
5890695
&
5891618
&
5891746
\\\hline
picojpeg&
5959857
&
5960273
&
5960519
&
18121177
&
18121508
&
18121971
&
10716870
&
10717508
&
10717870
&
9256370
&
9256782
&
9257295
\\\hline
pointer-chase&
20396094
&
20396748
&
20397401
&
29398389
&
29399709
&
29400284
&
27395878
&
27396035
&
27396808
&
27393724
&
27394596
&
27395145
\\\hline
qrduino&
6053379
&
6053754
&
6054146
&
16414628
&
16415002
&
16415377
&
9814649
&
9815144
&
9815556
&
9400335
&
9400552
&
9400927
\\\hline
st&
347356
&
347388
&
347426
&
1194276
&
1194313
&
1194350
&
735278
&
735316
&
735352
&
565223
&
565257
&
565301
\\\hline
st-opt&
285404
&
285450
&
285491
&
950749
&
950788
&
950817
&
563612
&
563645
&
563688
&
432199
&
432229
&
432272
\\\hline
statemate&
6160619
&
6160794
&
6161050
&
16283797
&
16283908
&
16284360
&
8143852
&
8144085
&
8144236
&
7680657
&
7680893
&
7682602
\\\hline
ud&
3302205
&
3302306
&
3302411
&
7836376
&
7836490
&
7836586
&
4417523
&
4417650
&
4417762
&
3438208
&
3438328
&
3438462
\\\hline

\end{tabular}
}
\caption{\label{tab:benchmark_dyn_t} Benchmark results for the dynamic target branch predictor (left) and the strictly no speculation (right) over 100 executions.}

\end{figure*}

\section{Raw Benchmark Results}
\label{appendix:rawBench}

\subsection{VexRiscv}
The
Table
shown
in
\cref{tab:benchmark_dyn_t})
lists
the
mean
result
of
each
benchmark
on
VexRiscv
as
well
as
the
upper
and
lower
quartiles
over
100
executions.

\FloatBarrier

\subsection{Gem5}
The
Gem5
simulation
behaves
deterministically
and
hence
produces
no
noise.
The
raw
benchmark
results
are
shown
in
\cref{tab:benchmark_raw_gem5}.

\begin{figure}[H]
\centering
\resizebox{\columnwidth}{!}{\begin{tabular}{ |p{.1\textwidth}||
P{.1\textwidth}|P{.1\textwidth}|P{.1\textwidth}|P{.1\textwidth}|
}
\hline
\textbf{Benchmark} & \textbf{BP} & \textbf{NoSpec} & \textbf{BB Info} & \textbf{BB Resched} \\
\hline\hline
Coremark&
\multicolumn{4}{c|}{Did
not
compile}
\\\hline
aha-mont
&
670306500
&
2987507500
&
1910602500
&
1744066500
\\\hline
aha-mont-opt
&
69956500
&
314984500
&
193341500
&
184647000
\\\hline
crc32
&
1482883000
&
6445296500
&
3659252000
&
2962620000
\\\hline
crc32-opt
&
695790500
&
2221779500
&
764247000
&
764237000
\\\hline
cubic
&
36000500
&
120518500
&
86233500
&
75575000
\\\hline
edn
&
1397500000
&
4843962500
&
2273180500
&
1863205000
\\\hline
huffbench
&
\multicolumn{4}{c|}{Did
not
run}
\\\hline
matmult-int
&
626205000
&
2233828500
&
726828500
&
670825500
\\\hline
minver
&
103290000
&
495655500
&
450553000
&
405312500
\\\hline
nbody
&
8393000
&
34323000
&
26280500
&
26177000
\\\hline
nettle-aes
&
1049202000
&
3874523500
&
1360225500
&
1243467000
\\\hline
nettle-sha
&
979399500
&
3521354500
&
1104238500
&
1086316500
\\\hline
picojpeg
&
1212690000
&
4543971500
&
3966359000
&
3528767000
\\\hline
pointer
chase
&
71722088000
&
76775731000
&
74809930500
&
74804141500
\\\hline
qrduino
&
1933871000
&
5771671000
&
3842855500
&
3706777000
\\\hline
st
&
31402500
&
110567000
&
58319500
&
45715000
\\\hline
st-opt
&
25673500
&
95303000
&
47212500
&
36370500
\\\hline
statemate
&
1175573000
&
4671848500
&
2352586500
&
2231463000
\\\hline
ud
&
513400500
&
2194281500
&
1189091500
&
1019462500
\\\hline
\end{tabular}
}
\caption{\label{tab:benchmark_raw_gem5} Raw benchmark results for BasicBlocker Gem5.}

\end{figure}
	
\end{appendices}

\end{document}